\documentclass[12pt,a4paper]{iopart}
\usepackage{hyperref}
\usepackage{cite}
\usepackage{graphicx}

\begin{document}

\title[Detection of $\pi^+\pi^-$atoms at DIRAC]%
{Detection of $\pi^+\pi^-$atoms \\
with the DIRAC spectrometer at CERN}


\author{B~Adeva$^{p}$,
L~Afanasyev$^{l}$,
M~Benayoun$^{e}$,
A~Benelli$^{r}$,
Z~Berka$^{b}$,
V~Brekhovskikh$^{o}$,
G~Caragheorgheopol$^{m}$,
T~Cechak$^{b}$,
M~Chiba$^{k}$,
S~Constantinescu$^{m}$,
S~Costantini$^{r}$,
C~Detraz$^{a}$,
D~Dreossi$^{g}$,
D~Drijard$^{a}$,
A~Dudarev$^{l}$,
I~Evangelou$^{d}$,
M~Ferro-Luzzi$^{a}$,
M~V~Gallas$^{p,a}$,
J~Gerndt$^{b}$,
R~Giacomich$^{g}$,
P~Gianotti$^{f}$,
D~Goldin$^{r}$,
F~G\'omez$^{p}$,
A~Gorin$^{o}$,
O~Gorchakov$^{l}$,
C~Guaraldo$^{f}$,
M~Hansroul$^{a}$,
R~Hosek$^{b}$,
M~Iliescu$^{f,m}$,
N~Kalinina$^{n}$,
V~Karpukhin$^{l}$,
J~Kluson$^{b}$,
M~Kobayashi$^{h}$,
P~Kokkas$^{d}$,
V~Komarov$^{l}$,
V~Kruglov$^{l}$,
L~Kruglova$^{l}$,
A~Kulikov$^{l}$,
A~Kuptsov$^{l}$,
I~Kurochkin$^{o}$,
K-I~Kuroda$^{l}$,
A~Lamberto$^{g}$,
A~Lanaro$^{a,f}$,
V~Lapshin$^{o}$,
R~Lednicky$^{c}$,
P~Leruste$^{e}$,
P~Levi Sandri$^{f}$,
A~Lopez Aguera$^{p}$,
V~Lucherini$^{f}$,
T~Maki$^{j}$,
N~Manthos$^{d}$,
I~Manuilov$^{o}$,
\fboxsep=2pt\fbox{L~Montanet$^a$},
J-L~Narjoux$^{e}$,
L~Nemenov$^{a,l}$,
M~Nikitin$^{l}$,
T~N\'u\~nez Pardo$^{p}$,
K~Okada$^{i}$,
V~Olchevskii$^{l}$,
A~Pazos$^{p}$,
M~Pentia$^{m}$,
A~Penzo$^{g}$,
J-M~Perreau$^{a}$,
C~Petrascu$^{f,m}$,
M~Pl\'o$^{p}$,
T~Ponta$^{m}$,
D~Pop$^{m}$,
G~F~Rappazzo$^{g}$,
A~Rodriguez Fernandez$^{p}$,
A~Romero$^{p}$,
A~Ryazantsev$^{o}$,
V~Rykalin$^{o}$,
C~Santamarina$^{p,q}$,
J~Saborido$^{p}$,
J~Schacher$^{q}$,
Ch~P~Schuetz$^{r}$,
A~Sidorov$^{o}$,
J~Smolik$^{c}$,
F~Takeutchi$^{i}$,
A~Tarasov$^{l}$,
L~Tauscher$^{r}$,
M~J~Tobar$^{p}$,
S~Trusov$^{n}$,
V~Utkin$^{l}$,
O~V\'azquez Doce$^{p}$,
P~V\'azquez$^{p}$,
S~Vlachos$^{r}$,
V~Yazkov$^{n}$,
Y~Yoshimura$^{h}$,
M~Zhabitsky$^{l}$ and
P~Zrelov$^{l}$}

\address{$^{a}$CERN, Geneva, Switzerland}
\address{$^{b}$Czech Technical University, Prague, Czech Republic}
\address{$^{c}$Institute of Physics ACSR, Prague, Czech Republic}
\address{$^{d}$Ioannina University, Ioannina, Greece}
\address{$^{e}$LPNHE des Universites Paris VI/VII, IN2P3-CNRS, France}
\address{$^{f}$INFN - Laboratori Nazionali di Frascati, Frascati, Italy}
\address{$^{g}$INFN - Trieste and Trieste University, Trieste, Italy}
\address{$^{h}$KEK, Tsukuba, Japan}
\address{$^{i}$Kyoto Sangyo University, Kyoto, Japan}
\address{$^{j}$UOEH-Kyushu, Japan}
\address{$^{k}$Tokyo Metropolitan University, Japan}
\address{$^{l}$JINR Dubna, Russia}
\address{$^{m}$IFIN-HH, National Institute for Physics and Nuclear Engineering, Bucharest, Romania}
\address{$^{n}$Skobeltsin Institute for Nuclear Physics of Moscow State University Moscow, Russia}
\address{$^{o}$IHEP Protvino, Russia}
\address{$^{p}$Santiago de Compostela University, Spain}
\address{$^{r}$Basel University, Switzerland}
\address{$^{q}$Bern University, Switzerland}

\ead{Leonid.Afanasev@cern.ch}


\begin{abstract}
  The goal of the DIRAC experiment at CERN is to measure with high
  precision the lifetime of the $\pi^+\pi^-$~atom ($A_{2\pi}$), which
  is of order $3\times10^{-15}$~s, and thus to determine the s-wave
  $\pi\pi$-scattering lengths difference $|a_{0}-a_{2}|$.
  $A_{2\pi}$~atoms are detected through the characteristic features of
  $\pi^+\pi^-$~pairs from the atom break-up (ionization) in the
  target. We report on a first high statistics atomic data sample
  obtained from p~Ni interactions at 24~GeV/$c$ proton momentum and
  present the methods to separate the signal from the background.
\end{abstract}

\submitto{\JPG}


\maketitle

\rightline{\textit{Dedicated to the memory of Lucien Montanet}}


\section{Introduction}\label{introduction}

The measurement of the lifetime of the $\pi^+\pi^-$~atom $A_{2\pi}$
\cite{Adeva95}, which is essentially determined by the
$\pi^+\pi^-\rightarrow \pi^{0}\pi^{0}$ reaction,
enables the determination of the combination $|a_0-a_2|$ of the s-wave
$\pi\pi$-scattering lengths for isopins $I=0$ and $2$ in a model
independent way
\cite{Deser54,Uretsky61,Bilenky69,Jallouli98,Ivanov98,Gasser01,Gashi02}
according to \cite{Gasser01}
\begin{equation}
   \frac{1}{\tau_{1s}}=
   \frac{2}{9} \; \alpha^3 \; p\left|a_0-a_2\right|^2 (1+\delta),
        \label{eq:gasser}
\end{equation}
\noindent with $\tau_{1s}$ the lifetime of the atomic ground state, $\alpha$
the electromagnetic coupling constant, $p$ the $\pi^{0}$ momentum in the
atomic c.m.s., and $\delta=(5.8\pm1.2)\times10^{-2}$ a correction due
to QED and QCD \cite{Gasser01}.

The $\pi\pi$ scattering lengths $a_0$ and $a_2$ have been
calculated within the framework of Chiral Perturbation Theory
\cite{Weinb79} by means of an effective Lagrangian with a 
precision at the percent
level \cite{Colan01NP}. The lifetime of $A_{2\pi}$ in the ground state
is predicted to be $\tau_{1s}=(2.9\pm0.1)\times10^{-15}$~s.  These
results are based on the assumption that the spontaneous chiral
symmetry breaking is due to a strong quark condensate as recently
suggested \cite{Colan01PRL,Pislak01}.  An alternative scenario with an
arbitrary value of the quark condensate \cite{Knecht95} allows for
larger $a_0$, $a_2$ compared with those of the standard scheme
\cite{Colan01NP}. A measurement of the scattering lengths will thus
contribute crucially to the current understanding of chiral symmetry
breaking in QCD and constrain the magnitude of the quark condensate.

The differential production cross section of $A_{2\pi}$ atoms can be
obtained from the double differential two-pion production cross section
($\sigma_0$) without a Coulomb final state interaction
\cite{Nem85}:
\begin{equation}\label{eq:prod}
\frac{d \sigma^A_n}{d{\vec{p}}_A}  =
(2\pi)^3 \frac{E_A}{M_A} \; | \Psi_n (0) |^2
\left.\frac{d\sigma_0}{d\vec{p}_1 d\vec{p}_2}
\right|_{\vec{p}_1=\vec{p}_2}
\;,
\end{equation}
where the differential production cross section $\sigma^A_n$ for atoms
with principal quantum number $n$ and zero angular momentum
depends on mass ($M_A$), momentum ($\vec{p}_A$) and  energy ($E_A$) of the
atom in the lab frame, and on the square of the Coulomb atomic wave
function at zero distance $| \Psi_n (0) |^ 2$. The laboratory momenta
of the $\pi^+$ and $\pi^-$ are denoted by $\vec{p}_1,~\vec{p}_2$,
respectively.  On the basis of Eq.\,\ref{eq:prod} and using the
Fritiof~6 generator, yields for $A_{2\pi}$ in proton nucleus
interactions have been calculated as a function of their energy and
angle in the proton energy range from 24~GeV to 1000~GeV
\cite{Nem85,Gorch96,Gorch00}. For a Ni target and a 24\,GeV/c proton
beam, $\sim 7\times 10^{-7}$ atoms are produced per proton
interaction, of which, however, only $\sim 1\times 10^{-9}$ are 
detectable in the experiment, due to momentum and angular acceptance
of the DIRAC apparatus.

\begin{figure}[htb]
\begin{minipage}[t]{0.49\textwidth}
\centering
  \includegraphics[width=0.86\textwidth]{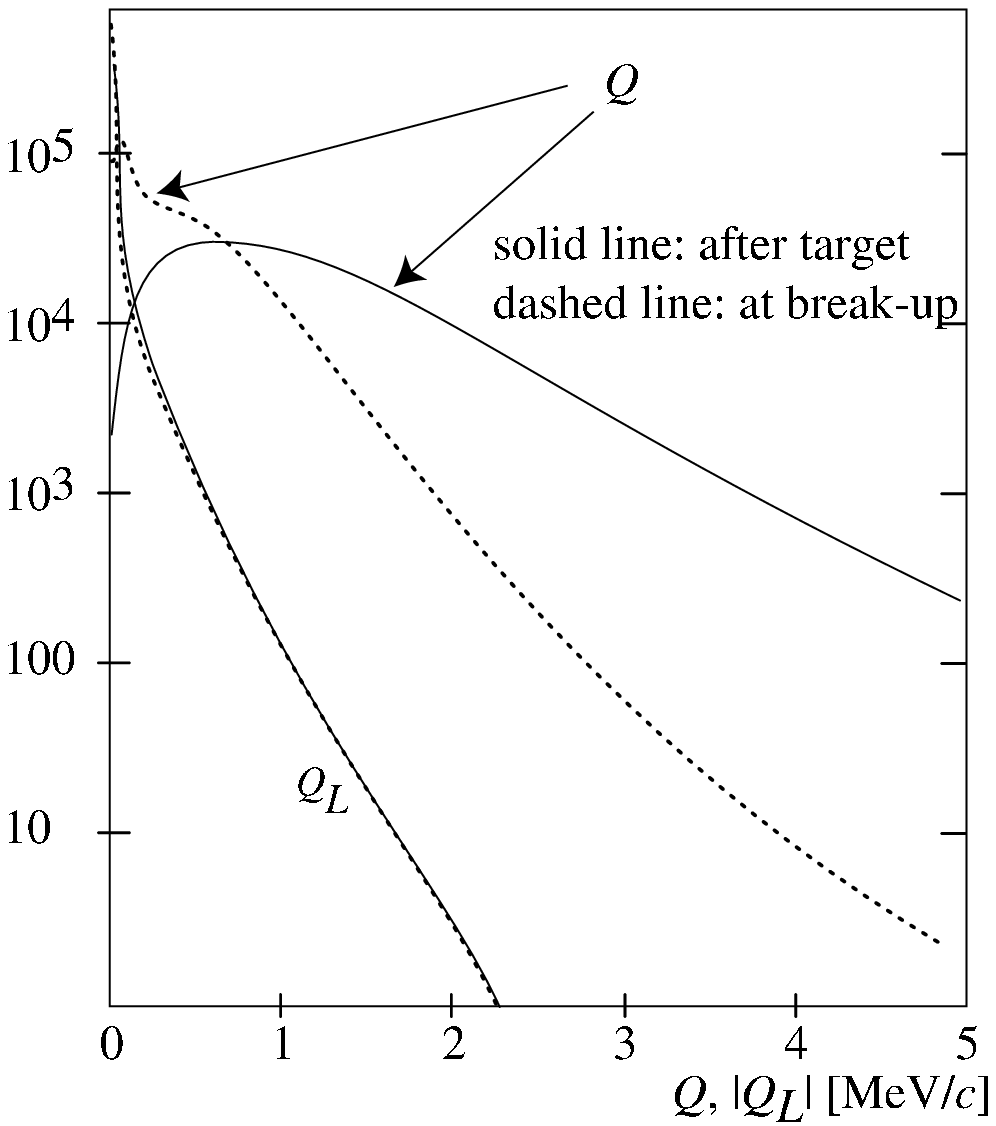}
\caption{Relative momentum distributions ($Q$, $Q_{L}$) for atomic
  $\pi^{+}\pi^{-}$ pairs at break-up and at the exit of the target
  (event generator, no spectrometer simulation).  Note that $Q_{L}$ is
  almost not affected by multiple scattering in the target.}
\label{fig:At}
\end{minipage}
\hfill
\begin{minipage}[t]{0.49\textwidth}
\centering
\includegraphics[width=0.9\textwidth]{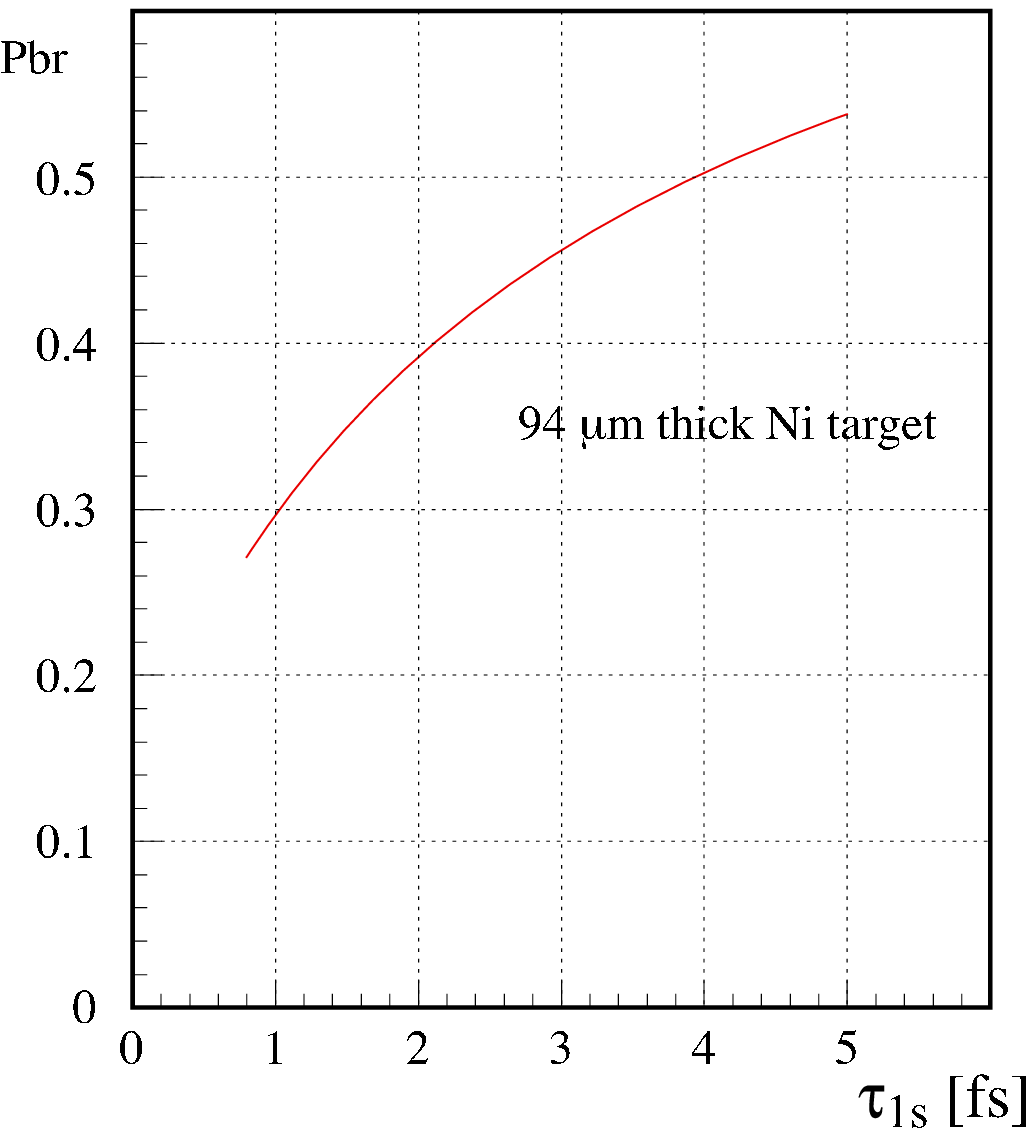}
\caption{Break-up probability $Pbr$ as a function of the ground state
  lifetime $\tau_{1s}$ of the $A_{2\pi}$ atom for a Ni-target and
  DIRAC conditions.}
\label{fig:Pbr-tau}
\end{minipage}
\end{figure}

The method for observing $A_{2\pi}$ and measuring its lifetime has
been proposed in \cite{Nem85}.  Pairs of $\pi^+\pi^-$ are produced
mainly as unbound ("free") pairs, but sometimes also as $A_{2\pi}$.
The latter may either decay, get excited or break up into $\pi^+\pi^-$
pairs (atomic pairs) after interacting with target atoms.  Due to
their specific kinematical features, these atomic pairs are
experimentally observable.  For thin targets ($10^{-3}X_0$) the
observable relative momentum $Q$ in the atomic pair system is
$Q\leq3$\,MeV/$c$. Their yield is $\sim10\%\div20\%$ of the number of
free pairs in the same $Q$ interval.  In Fig.\,\ref{fig:At} the
relative momentum distributions of atomic pairs are shown for $Q$ and
$Q_{L}$ (relative momentum component along the flight direction of the
pair) at the moment of break-up and at the exit of the target, as
obtained with an event generator based on \cite{Schum02}.  The number
of atomic pairs is a function of the atom momentum and depends on the
dynamics of the $A_{2\pi}$ interaction with the target atoms and on
the $A_{2\pi}$ lifetime \cite{Afan96}.  The theory of the $A_{2\pi}$
interaction with ordinary atoms allows the calculate of the relevant
cross sections
\cite{Schum02,Afan96,Dulian83,Mrowc,Halab99,Taras91,Voskr98,Afan99,Ivanov99gl,Heim00,Heim01,Afan02}.
For a given target thickness, the theoretical breakup probability for
$A_{2\pi}$ is precise at the 1\% level and uniquely linked to the
lifetime of the atom \cite{Schum02,Santa03}.  In
Fig.\,\ref{fig:Pbr-tau} the break-up probability as a function of the
lifetime is displayed for a 94~$\mu$m thick Ni target and for atomic
pairs accepted in the DIRAC spectrometer.

The first observation of $A_{2\pi}$ \cite{Afan93} has been achieved in
the interaction of 70~GeV/$c$ protons with Tantalum at the Serpukhov
U-70 synchrotron.  In that experiment, the atoms were produced in an
$8\,\mu$m thick Ta target, inserted into the internal proton beam.
With only $270\pm50$ observed atomic pairs, it was already possible to
set a lower limit on the $A_{2\pi}$ lifetime \cite{Afan94,Afana96}:
$\tau > 1.8\times10^{-15}$~s $(90\%~CL)$.

In this paper, we present the first high statistics experimental data
on $A_{2\pi}$ production on a Ni target at an external proton beam of
the CERN PS and demonstrate the feasibility of the lifetime
measurement.  We are not attempting to deduce a lifetime here, as this
requires a highly involved analysis of the normalization when
evaluating the break-up probability.


\section{The DIRAC experimental setup}\label{setup}

\begin{figure}[ht]
\begin{center}
\includegraphics[width=0.8\textwidth]{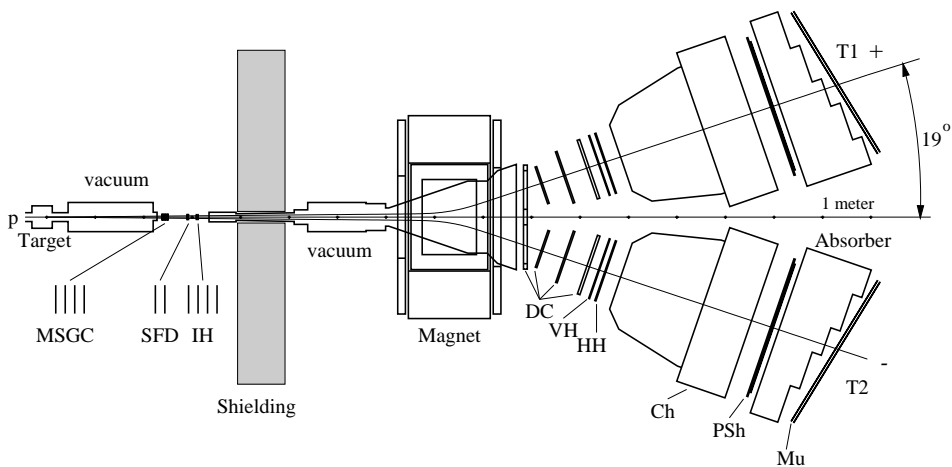}
\caption
{ Schematic top view of the DIRAC spectrometer.  Upstream of the
  magnet: target, microstrip gas chambers (MSGC), scintillating fiber
  detectors (SFD), ionization hodoscopes (IH) and iron shielding.
  Downstream of the magnet: drift
  chambers (DC), vertical and horizontal scintillation hodoscopes (VH,
  HH), gas Cherenkov counter (Ch), pre-shower detectors (PSh) and,
  behind the iron absorber, muon detectors (Mu).}
\label{fig:schem}
\end{center}
\end{figure}

The DIRAC setup is designed to detect oppositely charged pion pairs of
low relative c.m. momenta with high resolution using a magnetic double
arm spectrometer at a 24~GeV/$c$ extracted proton beam of the CERN PS
and an especially low material budget in the secondary particle path. The
spectrometer setup is shown in Fig.\,\ref{fig:schem}. A detailed
description may be found in \cite{setup}.

The proton beam intensity during data taking was $0.9\times 10^{11}$
per spill with a spill duration of $400\div 450$\,ms. The beam line
was designed such as to keep the beam halo negligible
\cite{PS/CA/Note97-16}. The horizontal and vertical widths of the beam
spot were $\sigma_{\mathrm{hor}}= (0.80\pm 0.08)$~mm and
$\sigma_{\mathrm{vert}}= (1.60\pm 0.07)$~mm, respectively
\cite{note02-01}.  The targets we report here were 94 and 98\,$\mu$m
thick Ni foils, corresponding to $\sim 6.4\times 10^{-4}$ nuclear
interaction probability \cite{PDG2002} or $6.7\times 10^{-3}$
radiation length. The transverse dimensions of the circular targets
(4.4\,cm diameter) were sufficient to contain the proton beam fully 
and to exclude possible interactions of beam halo with the target holder.

The spectrometer axis is inclined upward by $5.7^\circ$ with respect
to the proton beam. Particles produced in the target propagate in
vacuum up to the upstream (with respect to the magnet) detectors and
then enter into a vacuum chamber which ends at the exit of the
magnet. The exit window is made of a 0.68\,mm thick Al foil. The
angular aperture is defined by a square collimator and is 1.2\,msr
($\pm1^\circ$ in both directions).  The dipole magnet properties are
$B=1.65$~T and $BL=2.2$~Tm.

The upstream detectors are 2.5 to 3\,m away from the target and cover
an area of roughly $10\times 10$\,cm$^2$.  The microstrip gas chambers
(MSGC) consist of four planes: X, Y, U and V, with rotation angles of
$0^\circ$, $90^\circ$, $5^\circ$, $85^\circ$ with respect to the
X-plane. Each plane has 512 anode strips with a pitch of 0.2\,mm.
Clustering results in a spatial resolution for single tracks of
54\,$\mu$m.  The two planes (X, Y) of the scintillating fiber detector
(SFD) provide both coordinate and timing information. Each plane
contains 240 fiber columns (0.44\,mm pitch), each column consisting of
5 fibers of 0.5~mm diameter. They are read out through multichannel
position-sensitive photomultipliers and an analog signal processor
that produces information for the appropriate TDC channel. The space
and time resolutions are 130\,$\mu$m (rms) and 0.8\,ns, respectively.
The analog signal processor merges two adjacent hits into one,
depending on the relative pulse hight of the two signals and on their
time difference. No merging takes place if the time difference is
larger than 5\,ns.
The ionization hodoscopes (IH) serve for fast triggering and
identification of unresolved double tracks through an energy loss
measurement. The IH detector is composed of two vertical (X) and two
horizontal (Y) layers, each with 16 slabs of plastic scintillator
($1\times 7 \times 110$\,mm$^{3}$). The read-out provides logic
(TDCs, trigger processors) and analog (ADCs) information.  The
total thickness of all upstream detectors, including the vacuum
channel windows, is $6.0\times10^{-2}X_0$.

The two arms of the spectrometer are identically equipped. Four sets
of drift chambers are used (DC1 to DC4).  DC1 and DC4 have two X and
two Y planes each, DC1 has in addition two inclined W planes
(11.3$^{\circ}$ with respect to the X-wires).  DC2 and DC3 have one X
and one Y plane each.  The sensitive areas range from
$0.8\times0.4$\,m$^2$ to $1.28\times0.4$\,m$^2$, the signal wire pitch
is 10\,mm. The space resolution is better than 90\,$\mu$m.  The
vertical hodoscopes (VH) supply time-of-flight information and serve
trigger purposes. The hodoscope is made of 18 plastic scintillation
counters ($2.2 \times 7 \times 40$\,cm$^{3}$).  The time resolution is
127\,ps.  The horizontal hodoscope (HH) consists of 16 plastic
scintillators ($2.5\times2.5\times130$\,cm$^3$).  It serves
essentially trigger purposes.  The threshold Cherenkov detectors (Ch),
are used to identify electrons (positrons) and to reject pairs
containing an electron and/or a positron.  The radiator is Nitrogen
gas at normal pressure and ambient temperature.  The average number of
photoelectrons for particles with $\beta \approx$1 is larger than 16
and the efficiency more than 99.8\%.  Pion contamination above the
detection threshold is estimated to be less than 1.5\%.  The
pre-shower detector (Psh) consists of 8 elements, each comprising a Pb
converter and a scintillator. Off-line analysis of the amplitudes from
Psh provides additional $e$/$\pi$ separation. Each muon detector is
made of a thick iron absorber followed by two planes of
plastic-scintillator counters with 28 counters per plane.

The momentum range covered by the spectrometer is $1.2\div8$\,GeV/$c$.
The relative resolution on the lab-momentum is dominated by the
multiple scattering downstream of the vertical SFD detector and the
spatial resolution of the drift chambers. While the former leads to a
momentum independent resolution, the second leads to a slight increase
with momentum. Studies of the $\Lambda$ signal in $\pi^{-}$p track
pairs \cite{santiago-lambda} result in $\sigma_{p}/p =
2.8\times 10^{-3}$ and increasing to $3.3\times 10^{-3}$ at 8\,GeV/$c$.
The resolutions on the relative c.m.-momentum of $\pi^+\pi^-$ pairs
from atomic break-up, $\vec{Q}=(Q_{x}, Q_{y}, Q_{L})$, are in the
plane transverse to the total momentum $\vec{p}_{\pi^+\pi^-}^{~Lab}$,
$\sigma_{Q_x}\approx\sigma_{Q_y}\leq 0.49$\,MeV/$c$ and longitudinally
$\sigma_{Q_L}=0.50$\,MeV/$c$ \cite{dirac-resolution} \footnote{A
  momentum dependence of $\sim 0.016~[\mathrm{MeV}/c]\times
  p_{\pi^+\pi^-}^{~Lab}$ ($p$ in GeV/$c$) is included in these
  numbers.}.  The setup allows to identify electrons (positrons),
protons with $p\leq 4$~GeV/$c$ (by time-of-flight) and muons (cf.
Fig.\,23 in \cite{setup}). It cannot distinguish $\pi$- from
$K$-mesons, but kaons constitute a negligible contamination
\cite{Gorch00,ADEV02}.

The trigger system \cite{multilevel} reduces the event rate down to a
level acceptable for the data acquisition system.  The on-line event
selection keeps almost all events with $Q_{L}< 22$\,MeV/$c$,
$Q_{T}=\sqrt{Q_{x}^{2}+Q_{y}^{2}}< 10$\,MeV/$c$ and rejects events
with $Q \geq 15$\,MeV/$c$ progressively \cite{note03-04}. In the first
level trigger \cite{1stlevel} a coincidence of VH, HH and PSh signals
and anticoincidence with Ch signals in both arms is treated as a
pion-pair event. A condition on the HH of the positive and
negative arms, $|\mathrm{HH}_{+}-\mathrm{HH}_{-}| \leq 2$ slabs,
rejects events with $Q_y > 10$\,MeV/$c$.  Fast hardware processors
\cite{multilevel,T3,DNA,note00-13} are used to decrease the first
level trigger rate by a factor 5.5.  The trigger rate in standard
conditions was around 700 per spill. The trigger accepts events in a
time window $\pm$20\,ns with respect to the positive arm and thus
allows for collecting also accidental events.  The trigger
system provides parallel accumulation of events from several other
processes needed for calibration such as $e^+e^-$ pairs, $\Lambda
\rightarrow p \pi^-$ or $~\pi^{\pm} \pi^+ \pi^-$ final states.

The data acquisition \cite{DAQ,readout} accepted up to 2000 events per
spill, at spill intervals as short as one second.

Dead times were studied in a run with 30\% higher intensity than
normal and depend on average beam intensity as well as on micro duty
cycle of the beam. An overall dead time of 33\% was found for triggers
and data acquisition, with an additional 10\% due to front-end
electronics \cite{deadtime}. At normal intensities, dead times are
lower. Biases due to dead times could not be found. When selecting
data offline for further analysis, run periods, have been eliminated
where problems with detectors or spill structure or micro duty cycle
were found.

The full setup, including detectors, detector responses, read-outs,
triggers and the magnet has been implemented into the detector simulation
DIRAC-GEANT \cite{GEANT-DIRAC} and into the DIRAC analysis package
ARIANE \cite{ARIANE-DIRAC} such that simulated data can be treated in
the same way as real measurements.


\section{Track reconstruction}\label{tracking}

Events of interest consist of two particles of opposite charge with
very low Q, resulting in two close lying tracks upstream that are
separated downstream into the two arms of the spectrometer.

The magnetic field was carefully mapped and fine-tuned to a relative
precision of $10^{-4}$. The final field map is used in the Monte Carlo
GEANT simulation of the experiment \cite{GEANT-DIRAC}. For tracking,
the field is summarized by a transfer function that uses as input the
position, direction and momentum of a track in a reference plane at
the exit of the magnet and calculates the position and direction of
that track in a reference plane five meters upstream of the magnet.
The transfer function consists of 4 polynomials with 5 parameters
each.

\begin{figure}[htb]
\begin{minipage}[t]{0.49\textwidth}
\centering
  \includegraphics[width=0.80\textwidth]{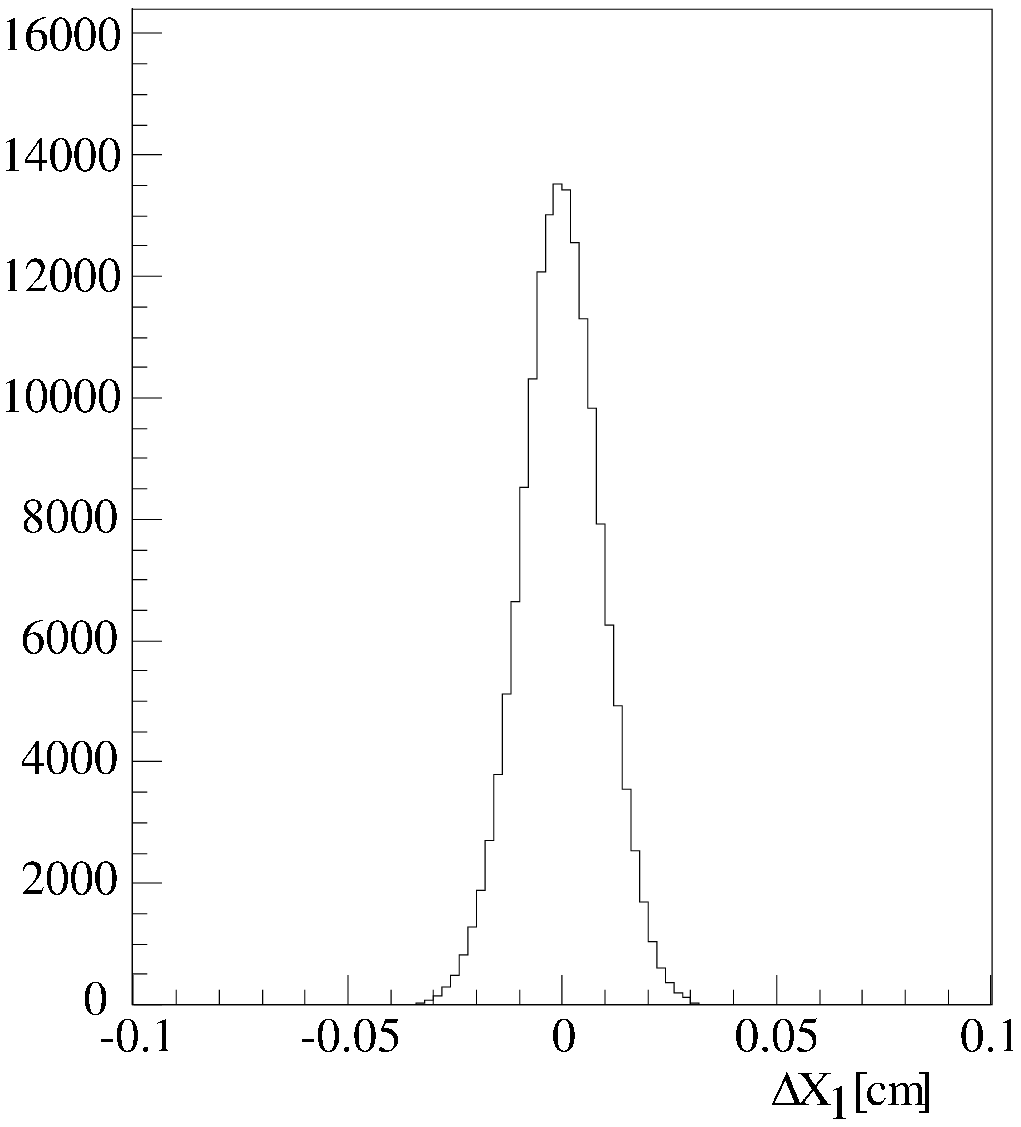}
\caption{ Difference between a hit coordinate
measured by the first X-plane of DC in the right arm and a coordinate
calculated with track parameters.}
\label{fig:DC1-residuals}
\end{minipage}
\hfill
\begin{minipage}[t]{0.49\textwidth}
\centering
\includegraphics[width=0.8\textwidth]{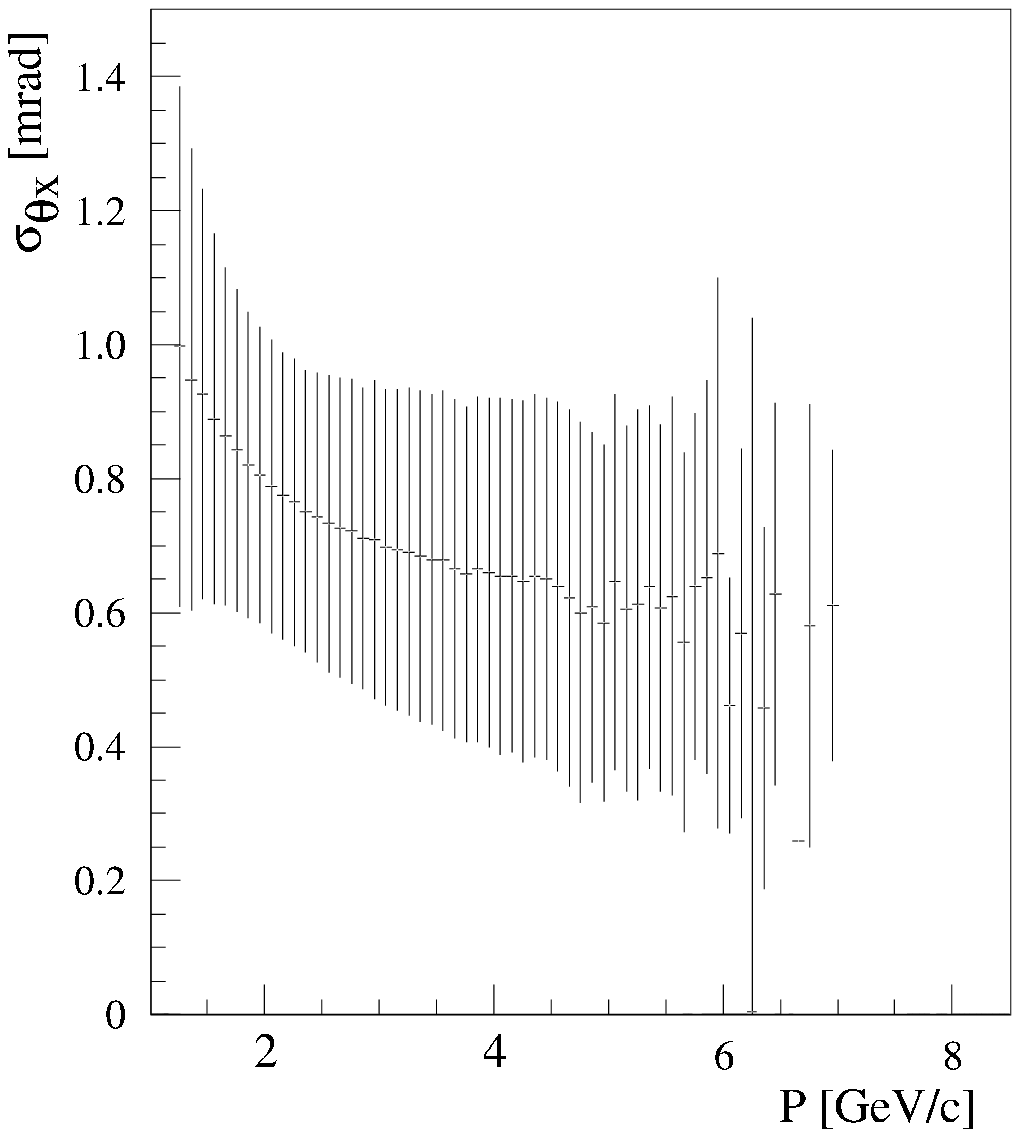}
\caption{Resolution for angles of DC-tracks in X projections}
\label{fig:ang-res}
\end{minipage}
\end{figure}

Track reconstruction starts from the downstream part.  A track
candidate is searched for in the horizontal and vertical planes of the
drift chambers separately. A "horizontal" candidate must have a hit
wire in one of the horizontal planes of the first (DC1) and last (DC4)
chamber, as well as corresponding hits in the HH-detectors. They
define a straight line. Hits close to that line are looked for in the
other four horizontal planes of DC1 to DC4.  Based on an acceptance
window, a candidate is accepted if at least two more hits are found.
The "vertical" candidates are found analogously. Finally, candidates
from horizontal and vertical wires are matched using the inclined
planes.  Projecting the spatial track parameters by means of the
magnetic field polynomials to the center of the beam spot at the
target provides a rough momentum estimate for each track candidate.
Using this estimate and the precise position measurements deduced from
drift-times, an overall track-fit for each candidate is made using a
standard least-squares method, where the full error matrix is used,
including the correlations (off-diagonal terms) induced by multiple
scattering.  Candidates are retained on the basis of a confidence
level cut. The downstream track candidates are thus defined spatially
with high precision.  This is demonstrated in
Fig.\,\ref{fig:DC1-residuals}, where the difference between the
measured hit coordinate in the first DC1 X-plane and the coordinate
obtained with the track parameters at the same DC1 plane is shown. The
spatial and angular resolutions depend on the particle momentum. They
show a $1/p$ dependence and level off at 6\,GeV/$c$. As an example the
angular resolution in X direction is shown in Fig.\,\ref{fig:ang-res}.
The one in the Y direction is about 4\% better.

The multiple scattering in the upstream detectors is such that in
practice they have to be considered as the effective source of each
track, and not the target as supposed above.
Therefore, each of the above tracks, projected through the magnet onto
the beam spot defines a reference line.  Along this line, hit
candidates are searched for in the upstream detectors within spatial
windows defined mainly by multiple scattering and within time windows
defined by the time-of-flight from the upstream detectors to the
vertical hodoscopes downstream and their time resolutions.

Using all identified upstream information, a track-fit is made for
each track-candidate by means of the Kalman-filter method
\cite{KALMAN}, starting from the first downstream hit and ending at
the exit window of the vacuum tube upstream of the MSGC detector.

With the hypothesis that the track originates from the target, the
intersection of the proton beam with the target provides another
measurement point for the Kalman filter, whose uncertainty is given by
the measured intensity distribution of the proton beam across the
target (cf. section \ref{setup}) \footnote{The beam position is
continuously monitored and calibrated during data taking. }. A cut on
the distance (15\,mm) of the track from the beam spot leads to the
final track selection \cite{BSTRACK}.

Alternatively, assuming that a track pair originates from the same
interaction, each pair candidate is fitted with the constraint that
both tracks intersect in the central plane of the target ("vertex
fit"). The complete $5\times 5$ uncertainty matrices of the two tracks
are used.  Tracks originating far from each other are rejected by a
threshold confidence level.
This procedure provides track parameters which are independent of the
precise knowledge of the beam position and beam width.
For the subsequent analysis these two procedures do not use the
MSGCs, for reasons of optimum efficiency.

Full tracking uses MSGC and SFD to connect all upstream hits by
straight lines.  These track candidates are then matched with the
downstream candidates. A vertex confidence level is calculated for
each track pair, corresponding to the hypothesis that both tracks
intersect at a common point lying on the target plane.
Correlations from the estimated errors from multiple scattering are
fully taken into account \cite{Pentia}. Both tracks are refitted at
the end, under the constraint of having a common vertex
\cite{S-TRACK}.

\begin{figure}[htb]
\begin{minipage}[t]{0.49\textwidth}
\centering
  \includegraphics[width=0.9\textwidth]{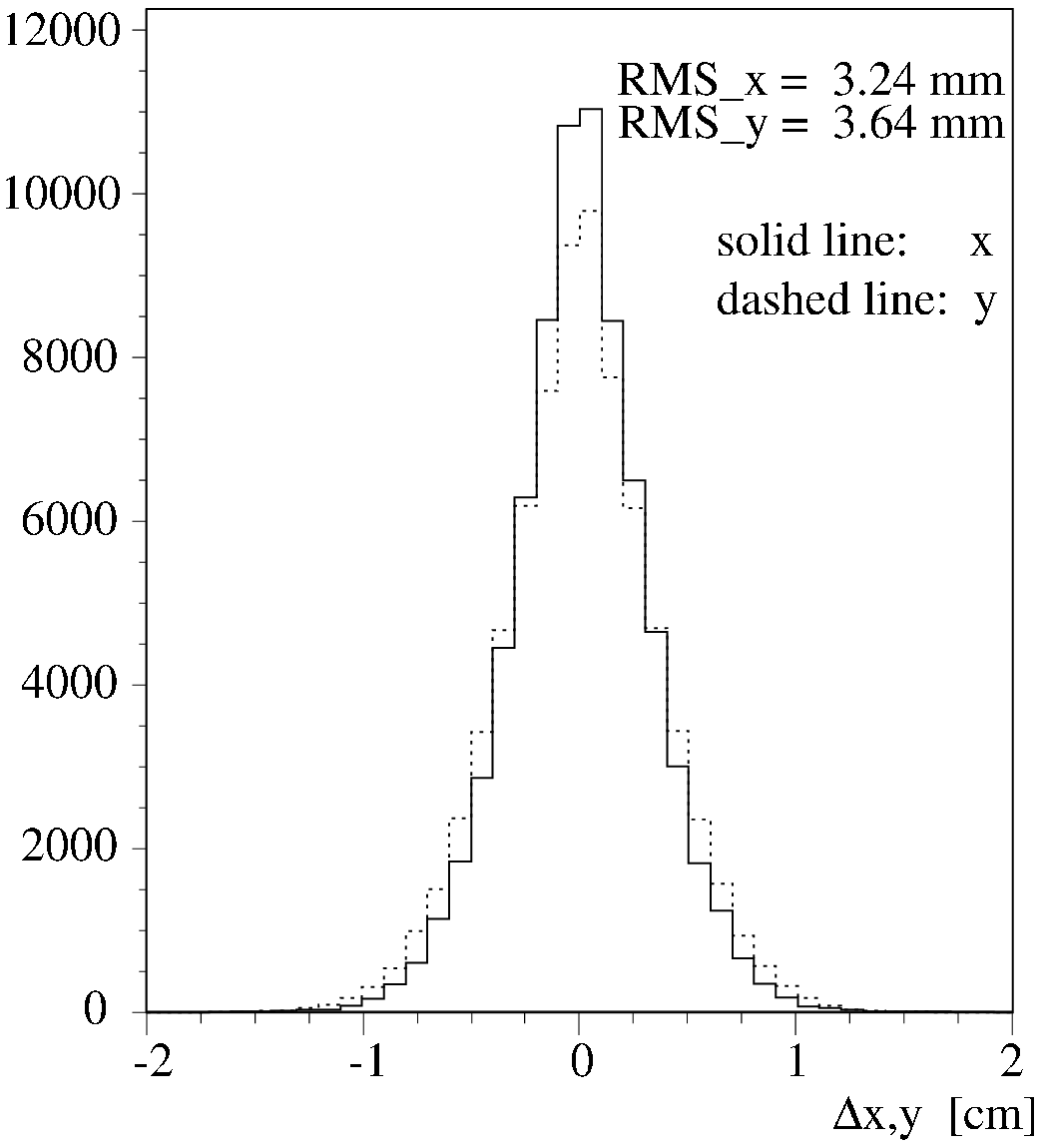}
\caption
{ Distances $\Delta x$ and $\Delta y$ between two reconstructed tracks at
  the target in the transverse plane.}
\label{fig:targ_delta}
\end{minipage}
\hfill
\begin{minipage}[t]{0.49\textwidth}
\centering
\centering
\includegraphics[width=0.87\textwidth]{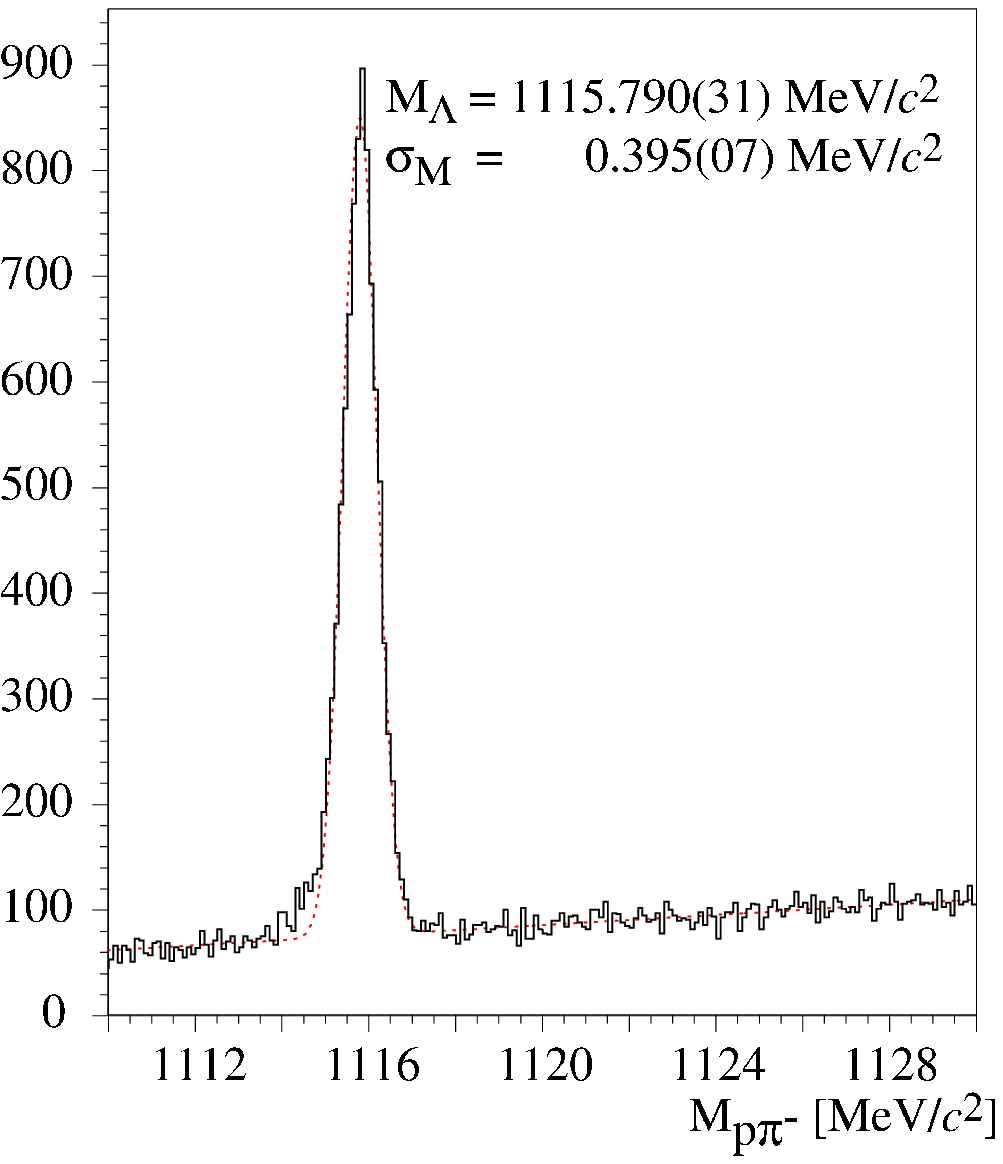}
\caption{ Invariant proton-pion mass ($M_{p\pi^{-}}$) distribution. 
The dashed line represents a fit of a Gaussian plus a straight line.}
\label{fig:lambda_mass}
\end{minipage}
\end{figure}

To illustrate that the reconstructed events do in fact originate from
the target, the distances between two tracks at the target are shown
in Fig.\,\ref{fig:targ_delta}. No vertex fit was done for these data.

The correctness of the alignment and magnetic field description were
verified by studying the decay of $\Lambda$ particles
\cite{kokkas-lambda}. Events with one downstream track per
spectrometer arm and a time-of-flight difference between the positive
particle and the negative one between 0 and 1.3\,ns were
selected.  Fig.\,\ref{fig:lambda_mass} shows, for a typical control
data sample, the invariant mass distribution from the two tracks under
the hypothesis proton-pion.  A fit of a Gaussian on a linear
background yields $M_{\Lambda}=(1115.790\pm0.031_{stat})$\,MeV/$c^{2}$
with $\sigma_{M_{\Lambda}}=(0.395\pm0.007_{stat})$\,MeV/$c^{2}$. The
measured width is entirely due to track reconstruction. The difference
($M_{p \pi^-}-M_{\Lambda}(PDG$~\cite{PDG2002})) in units of the
reconstruction error provides a measure of the correctness of the
error estimation. The resulting distribution is fitted by a Gaussian
with $\sigma=1.028\pm0.006$, showing that the reconstruction errors
are slightly underestimated.  The same distribution was studied as a
function of the pion momentum and found to be independent of it.  The
long term stability of the apparatus has been controlled using the
$\Lambda$ mass and the corresponding widths.


\section{Selection criteria}\label{selection}

An event is rejected if more than two downstream tracks in either of
the two arms are reconstructed. For each track the associated VH and
HH hodoscope slabs are required. In the case of two tracks in one arm,
the earlier in time is taken for further analysis. Events with more
than one track per arm constitute less than 4\% of the event sample.

Horizontal and vertical SFD hit candidates to be associated with a
track (see section \ref{tracking} for first momentum estimation) must
have times within a window of $\pm4$\,ns (corresponding to $\pm 5
\sigma$) with respect to the associated VH slab. Moreover, the hit
candidates must be found in a spatial window of $\pm (0.2+ 4.8/p
~[\mathrm{GeV/}c])$~cm (corresponding to $\pm 6\sigma$) with respect
to the point of intersection of the track candidate with the SFD
plane. This window is defined by multiple scattering in the
downstream material when backward-extrapolating the track. At least
one hit candidate is required. If there are more than four, the four
closest to the window center are kept.

After the first stage of Kalman filtering, only events are kept with
at least one track candidate per arm with a confidence level better
than 1\% and a distance to the beam spot in the target smaller than
1.5\,cm in x and y.  For such tracks, the SFD hits are redetermined in
a narrower ($\pm 1$cm) window around the new track intersect. Events
with less than four hits in the window per SFD plane and not more than
six in both planes are retained. Finally, all track pairs with
$|Q_x|<6$\,MeV/$c$, $|Q_y|<6$\,MeV/$c$ and $|Q_L|<45$\,MeV/$c$ are
selected for further analysis.

\begin{figure}[htb]
\begin{minipage}[t]{0.49\textwidth}
\centering
  \includegraphics[width=0.80\textwidth]{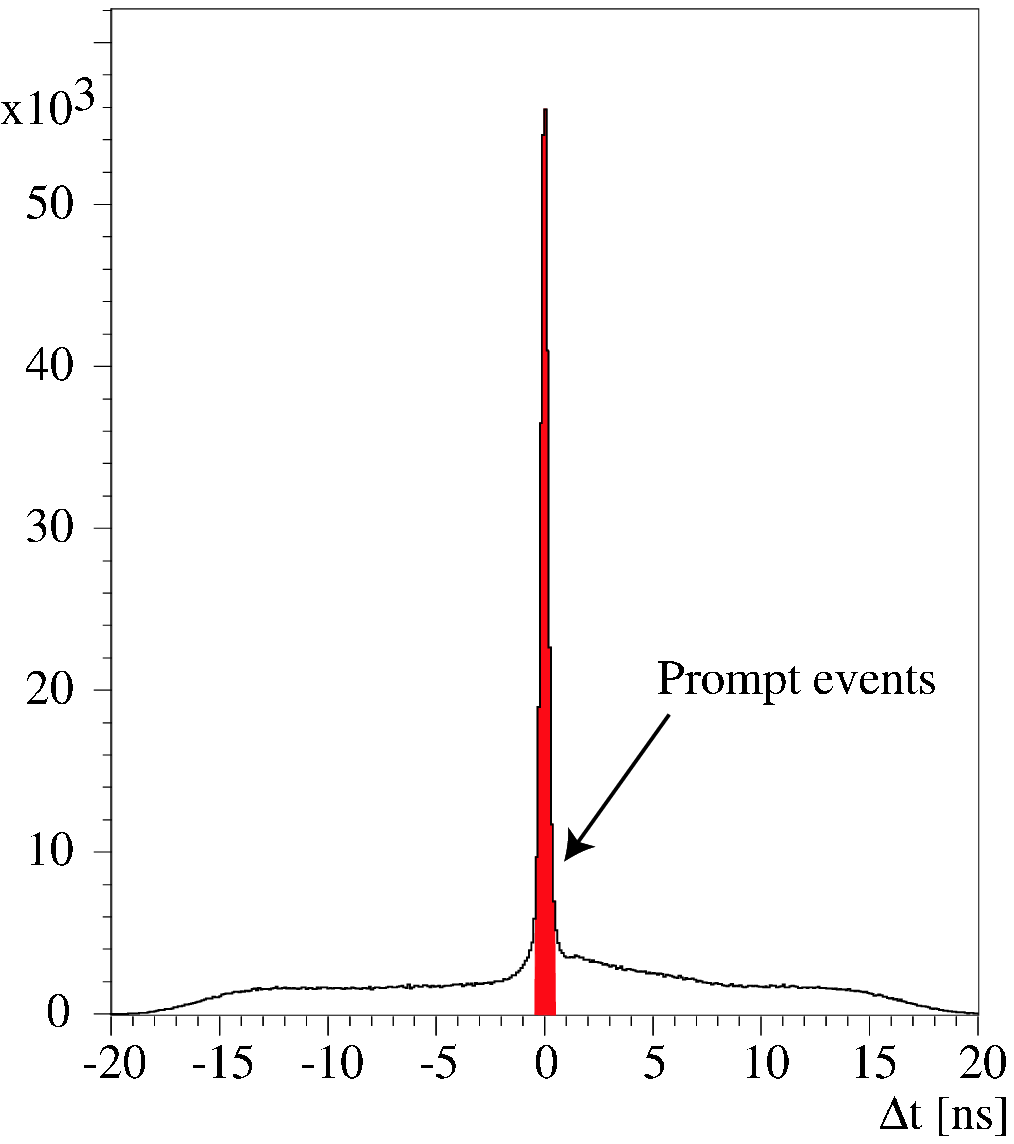}
\caption{Time difference between positive and negative vertical hodoscope
  slabs associated with the selected tracks. The asymmetry for
  positive differences is due to time correlated protons.}
\label{fig:delta-tVH}
\end{minipage}
\hfill
\begin{minipage}[t]{0.49\textwidth}
\centering
\includegraphics[width=0.83\textwidth]{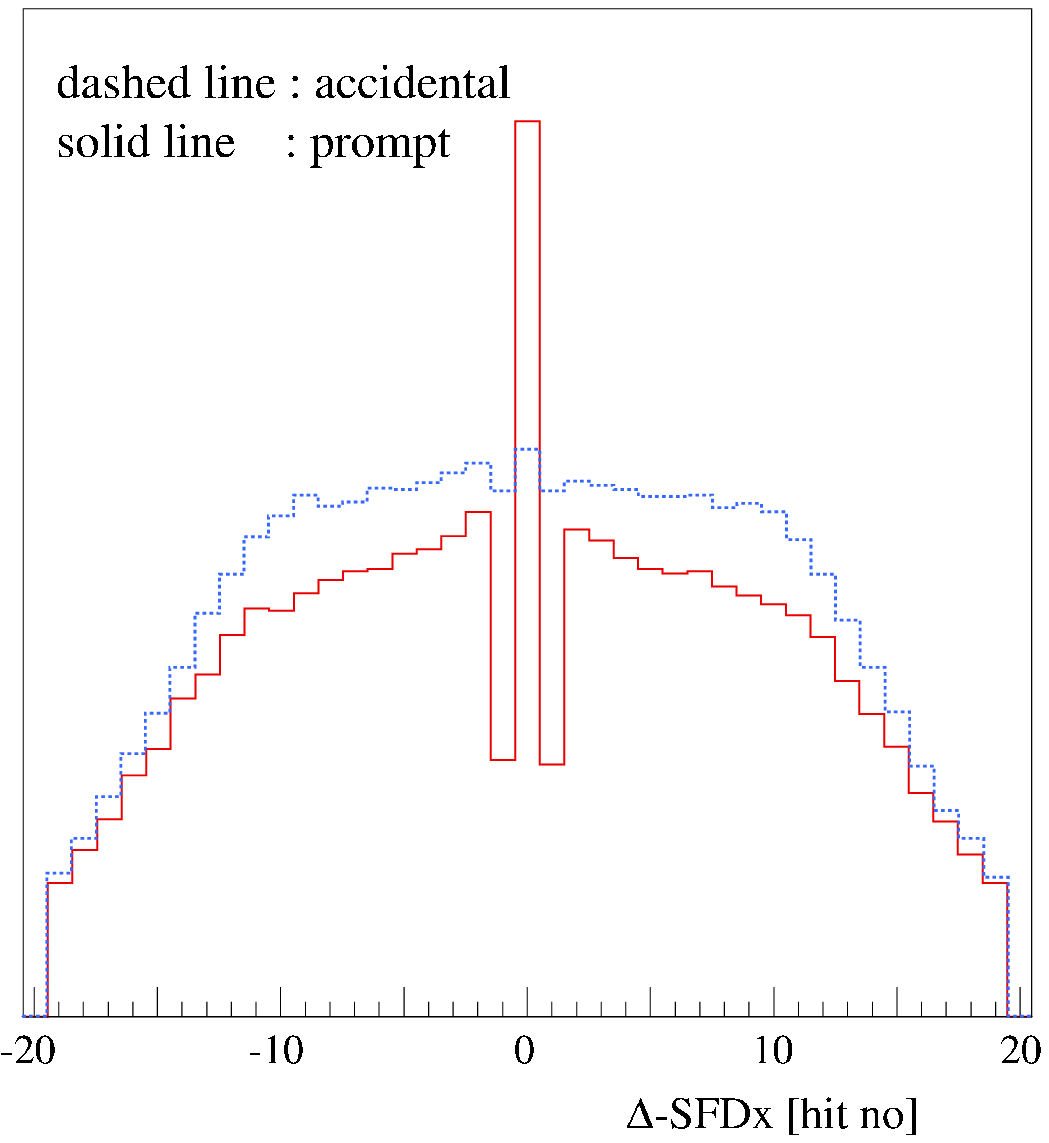}
\caption{Difference in hit number for selected SFD hits in the
  X-plane, for accidental and prompt events. Note the effect of
  merging of adjacent hits for prompt events. Arbitrary scale.}
\label{fig:DSFD}
\end{minipage}
\end{figure}

This preselection procedure retains 6.2\% of the $\pi \pi$ data
obtained with the Ni target in 2001. The time difference between the
positive and the negative arm, measured by the VHs, for events that
passed the preselection criteria is shown in
Fig.\,\ref{fig:delta-tVH}.

For the final analysis further cuts and conditions are applied:

\begin{itemize}

\item "prompt" events are defined by a time difference (corrected for
  the flight path assuming pions) measured by the VHs between the
  positive and the negative arm, $|\Delta t |\leq 0.5$\,ns,
  corresponding to a $\pm 2.7 \sigma$ cut.

\item "accidental" events are defined by time intervals $-15$~ns$ \leq
  \Delta t \leq -5$~ns and 7\,ns$ \leq \Delta t \leq 17$\,ns,
  conditioned by the read-out features of the SFD detector (cf.
  section \ref{setup}) and suppressing protons (cf.
  Fig.\,\ref{fig:delta-tVH}).

\item protons in prompt events are rejected by requiring momenta of
  the positive particle to be $p_{+}\leq 4$ GeV/$c$.

\item $e^{\pm}$ and $\mu^{\pm}$ are rejected through appropriate cuts
  on the Cherenkovs, the Pre-shower and the Muon counters
  \cite{note01-02}.

\item $Q_{T} \leq 4$\,MeV/$c$ and
  $|Q_L|<22$\,MeV/$c$.  The $Q_{T}$ cut preserves 98\% of the atomic
  signal, the $Q_{L}$ cut preserves background outside the signal
  region for defining the background below the signal.
  
\item the vertex-fit with highest confidence level from track and
  vertex fits is retained.

\item only events with at most two preselected hits per SFD plane are
  accepted. This provides the cleanest possible event pattern. This
  criterion is not used with the full tracking (cf. section
  \ref{tracking}).

\end{itemize}

\section{Signal and background in $\pi^+\pi^-$~atom detection}\label{background}

Once produced in proton-nucleus interactions, all $A_{2\pi}$ atoms
will annihilate if they propagate in vacuum.  In a finite target,
however, they interact electromagnetically with the target atoms, and
some of them break up.  The event generator for $\pi^+\pi^-$ pairs
from $A_{2\pi}$ break-up yielded $Q$ and $Q_{L}$ distributions shown
in Fig.\,\ref{fig:At} (cf. section \ref{introduction}). At the exit of
the target, multiple scattering has caused a considerable broadening
of the $Q$-distribution while the $Q_{L}$ distribution has remained
unchanged. The $Q_{L}$ distribution at break-up is already much
narrower than the $Q$-distribution because the break-up mechanism
itself affects mostly $Q_{T}$ (analogous to $\delta$ electron
production). These features led us to consider in the analysis not
only $Q$ but also $Q_{L}$ distributions .

The atomic pairs are accompanied by a large background.  The $Q$
distribution of all $\pi^+\pi^-$ pairs produced in single
proton-nucleus interactions in the target is described by:
\begin{equation}
   d N/d Q  = d N_{\rm C}/d Q + d N_{\rm nC}/d Q + d n_{A}/d Q.
    \label{eq:eqX}
\end{equation}
\noindent $N_{\rm C}$ is the number of $\pi^+\pi^-$ pairs
originating from short-lived sources (fragmentation, rescattering,
mesons and excited baryons that decay strongly). These pairs undergo
Coulomb interaction in the final state (Coulomb pairs). $N_{\rm nC}$
is the number of $\pi^+\pi^-$ pairs with at least one particle
originating from long-lived sources (mesons and baryons that decay
electromagnetically or weakly). They do not exhibit Coulomb final
state interaction (non-Coulomb pairs).  Finally, $n_A$ is the number
of $\pi^+\pi^-$ pairs from $A_{2\pi}$ break-up.

Accidental $\pi^+\pi^-$ pairs ($d N_{\rm acc}/d Q$) originate from
different proton-nucleus interactions and are uncorrelated in time,
i.e. they are neither affected by Coulomb nor by strong interaction in
the final state. Such events may also belong to the time window that
defines prompt events (see section \ref{selection}).

These backgrounds, as they are measured and reconstructed, are needed
for subtraction from the measured data in order to obtain the excess
produced by the atomic signal. The backgrounds may be obtained in
different ways.

One method is based on a Monte Carlo modelling of the background
shapes using special generators for the non-Coulomb (nC) and Coulomb
(C) backgrounds and accidental pairs \cite{note03-09}. Uniformity in
phase space is assumed for the backgrounds ($d N/d Q \sim Q^{2}$),
modified by the theoretical Coulomb correlation function $A_{\rm
  C}^{\rm theo}(Q)$ \cite{SAKH48} for C-background. Effects of the
strong final state interaction and the finite size of the pion
production region \cite{ll82} on the Coulomb correlation function are
small and neglected here. The differences of pion momentum
distributions for the different origins of $\pi^+\pi^-$ pairs are
implemented (cf. \cite{note01-01,note03-09}).  The generated events
are propagated through the detector (GEANT \cite{GEANT-DIRAC}).
Simulating the detectors, detector read-outs and triggers and
analyzing the events using the ARIANE analysis package of DIRAC
\cite{ARIANE-DIRAC} result in the high statistics distributions $d
N_{\rm C}/d Q$, $d N_{\rm nC}/d Q$ (cf. Eq.\,\ref{eq:eqX}) , and $d
N_{\rm acc}/d Q$. These distributions are used later to analyze the
measured distributions.  Though not necessary for the signal
extraction, the $A_{2\pi}$ signal was also simulated yielding $d
n_{A}/d Q$. Monte Carlo simulated distributions were
generated with about 10 times the statistics of the measured data.
The simulated backgrounds were obtained with dedicated generators,
without any additional tracks from the proton nucleus interaction. A
special Monte Carlo simulation with additional background tracks leads
essentially to a reduction in reconstruction efficiency.

The second method circumvents the Monte Carlo simulations of
background, detectors and triggers by relating the backgrounds to the
measured accidental background.  It was successfully applied in
\cite{Afan93}.  The assumption is that accidentals can be used to
describe the distribution of free $\pi^+\pi^-$ pairs, which then must
be corrected for final state interactions. By denoting $\Phi(Q) \equiv
d N_{\rm acc}^{meas}/d Q$, the experimental correlation function
$R(Q)$ is given by:

\begin{eqnarray}
R(Q)=\frac{1}{\Phi(Q)}
\left(
\frac{d N_{\rm nC}}{d Q}+\frac{d N_{\rm C}}{d Q}
\right)
=N \times [f + A_{\rm C}(Q)]
\label{eq:eqY}
\end{eqnarray}

Here $A_C(Q)$ is the Coulomb enhancement function~\cite{SAKH48}
smeared by multiple scattering in the target and the finite setup
resolution. In practice it can not be obtained by folding procedures
but must be obtained from Monte Carlo simulations described above. $N$
and $f$ are free parameters, and $\Phi(Q)$ is the measured spectrum of
accidentals corrected for the differences between prompt and
accidental momentum distributions \cite{note01-01}.

The measured accidental distributions had to be corrected for the
different recording conditions as compared to prompt events, such as
the SFD merging of adjacent hits, or track identification by time,
which is impossible for prompt events in case of ambiguities. In
Fig.\,\ref{fig:DSFD} the difference in selected SFD hits is shown for
accidental and prompt events for the X-plane (the Y-plane is
similar). The merging feature of the SFD read-out together with a
$\sim$ 5\% single track inefficiency of the SFD lead to the
enhancement for $\Delta$-SFD=0, and to the dips left and right of
it.The correction of the measured accidental spectra for prompt
conditions was done in two ways. On an event by event basis adjacent
hits of the measured accidentals were merged into one hit according to
the measured probability, and that hit was given the times of the two
original hits. Then tracking was started, and the timing conditions
could be applied as described above (cf. section \ref{selection}). An
other way of correcting was to construct accidental distributions from
uncorrected events, but giving up the timing conditions. The resulting
distributions were then given weights according to the measured
merging probabilities.


\section{Experimental data and atomic signal extraction}\label{signal}

\begin{figure}[htb]
\centering
\begin{tabular}{cc}
  \includegraphics[width=0.42\textwidth]{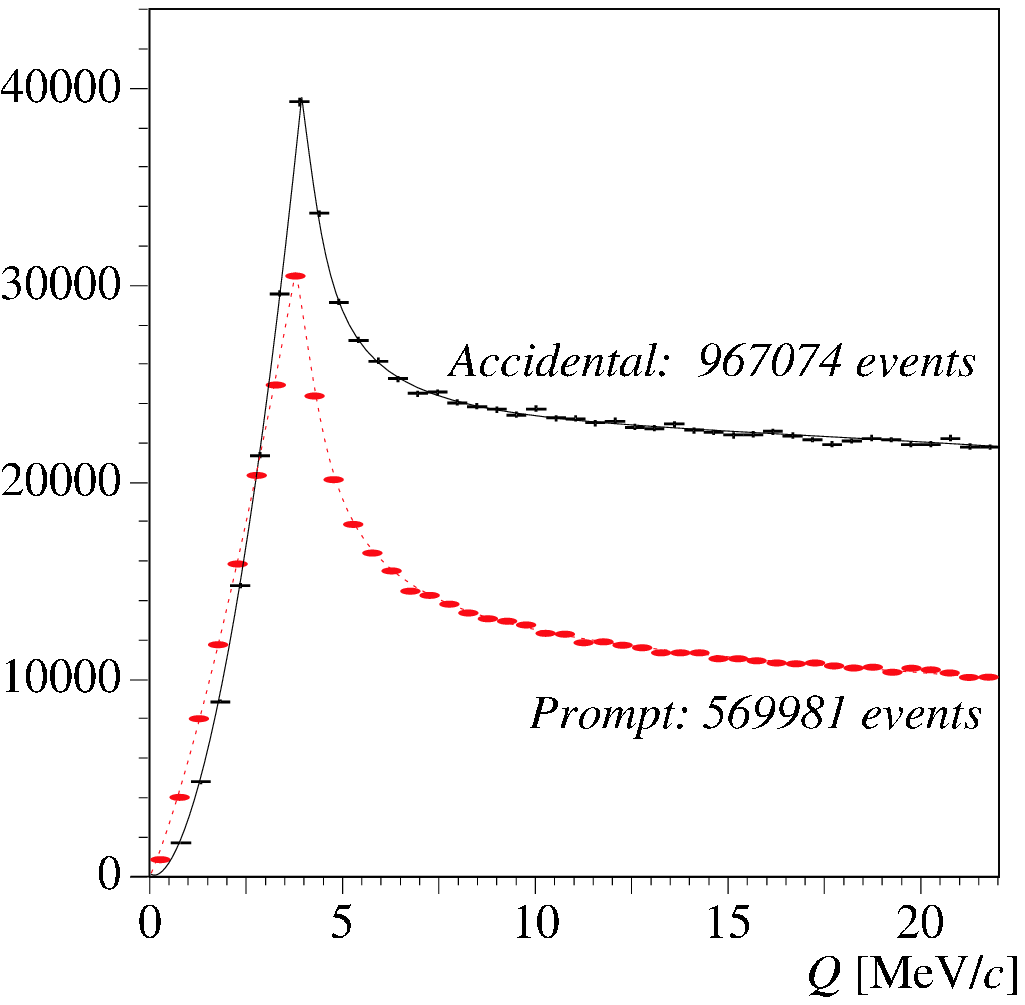}
  &
  \includegraphics[width=0.42\textwidth]{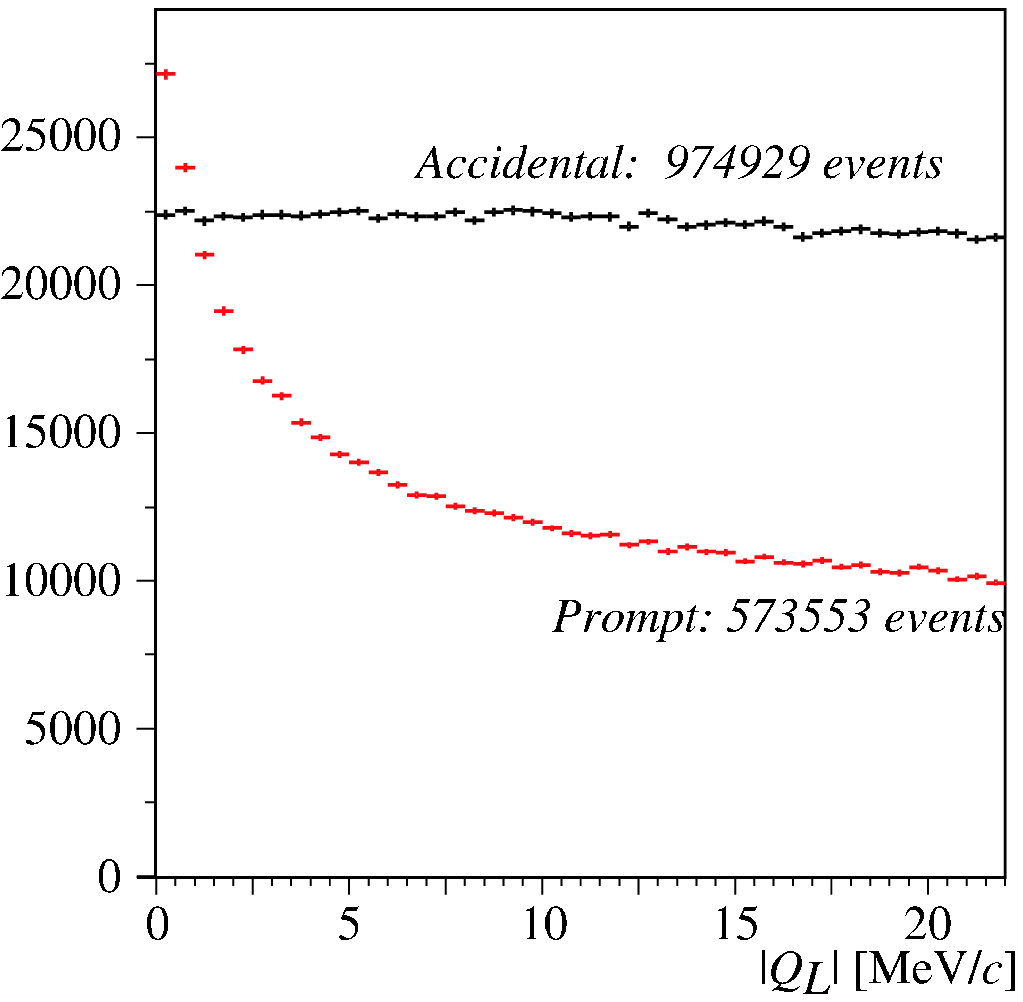}
\end{tabular}
\caption {Experimental $Q$ and $Q_{L}$ distributions for
prompt and accidental $\pi\pi$ events. Continuous lines serve as
guides for the eye. The sharp peak at $Q=4$\,MeV/$c$ is due to the cut
 $Q_{T}\leq 4$\,MeV/$c$. The number of events is given for the
displayed momentum window, the bin-width is 0.5\,MeV/$c$. Note the strong
Coulomb enhancement in the $Q_{L}$ distribution for prompt events.}
\label{fig:prompt-accid}
\end{figure}

DIRAC began data taking in autumn 1999.  Here we present data taken in
2001 using 94~$\mu$m and 98~$\mu$m thick Nickel targets. The
integrated proton flux through the target for these data is
$8.6\times10^{16}$, corresponding to $5.5\times10^{13}$ pNi
interactions and to $6.4\times10^{8}$ recorded $\pi\pi$ triggers. The
reconstructed accidental and prompt $\pi\pi$~pairs (see
Section~\ref{selection} for the cuts) are shown in
Fig.\,\ref{fig:prompt-accid}.

The Monte Carlo backgrounds or the correlation function $R(Q)$ (cf.
Section \ref{background}) are fitted to the measured prompt spectra in
$Q$ and $Q_{L}$ intervals which exclude the atomic signal (typically
$Q>4$\,MeV/$c$, $Q_{L}>2$\,MeV/$c$).  The amount of time-uncorrelated
events in the prompt region was determined to be 6.5\% of all prompt
events by extrapolating the accidental $\Delta t$ distribution to zero
and was subtracted from the prompt distributions (cf.
Fig.\,\ref{fig:delta-tVH}).  The fits provide the relative amounts of
non-Coulomb and Coulomb backgrounds and the parameters $N$ and $f$ of
Eq.\,\ref{eq:eqY}. As a constraint, these background components must be
the same for $Q$ and $Q_{L}$.

\begin{figure}[htb]
\centering
\begin{tabular}{cc}
   \includegraphics[width=0.47\textwidth]{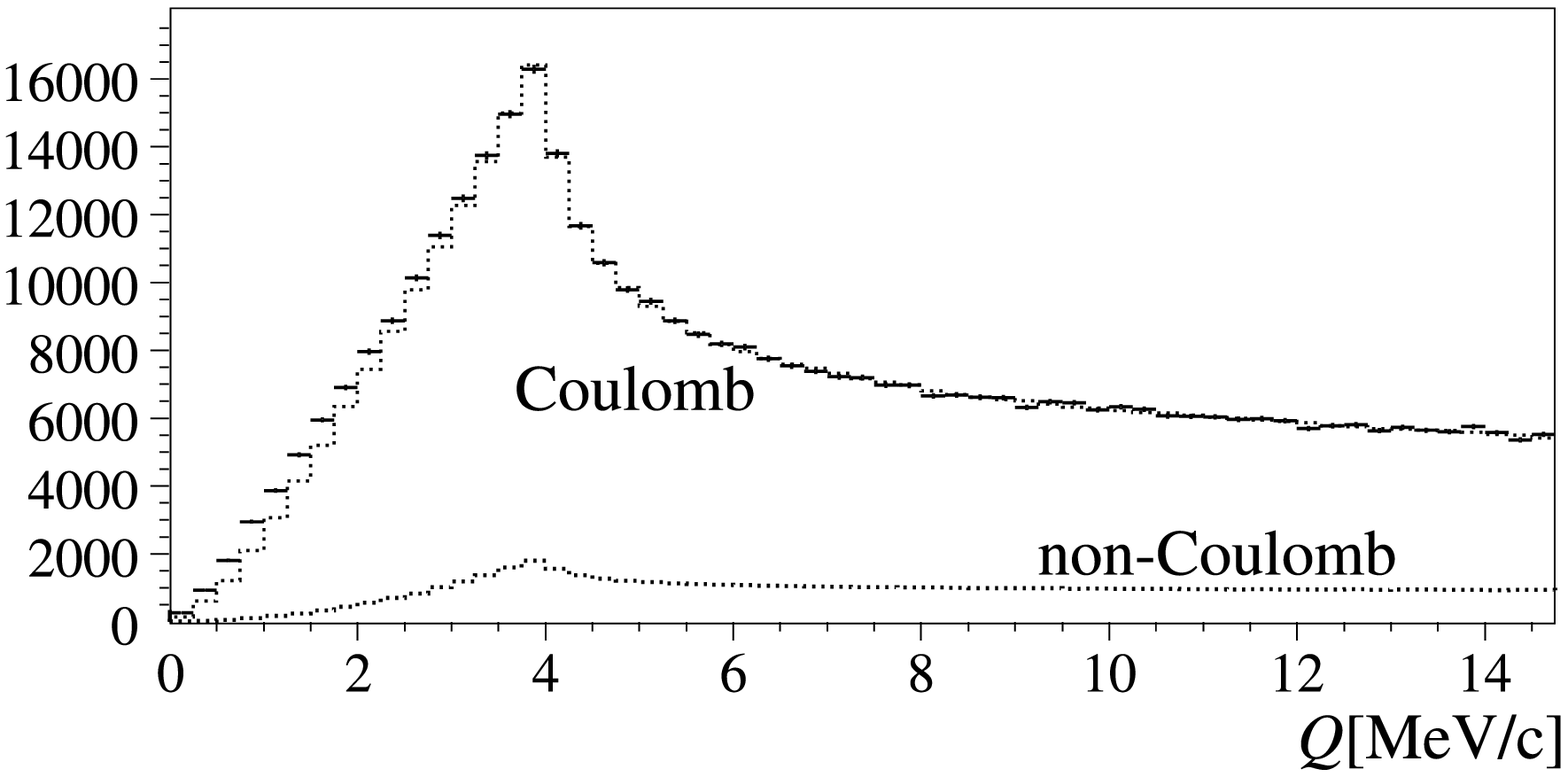}&
   \includegraphics[width=0.47\textwidth]{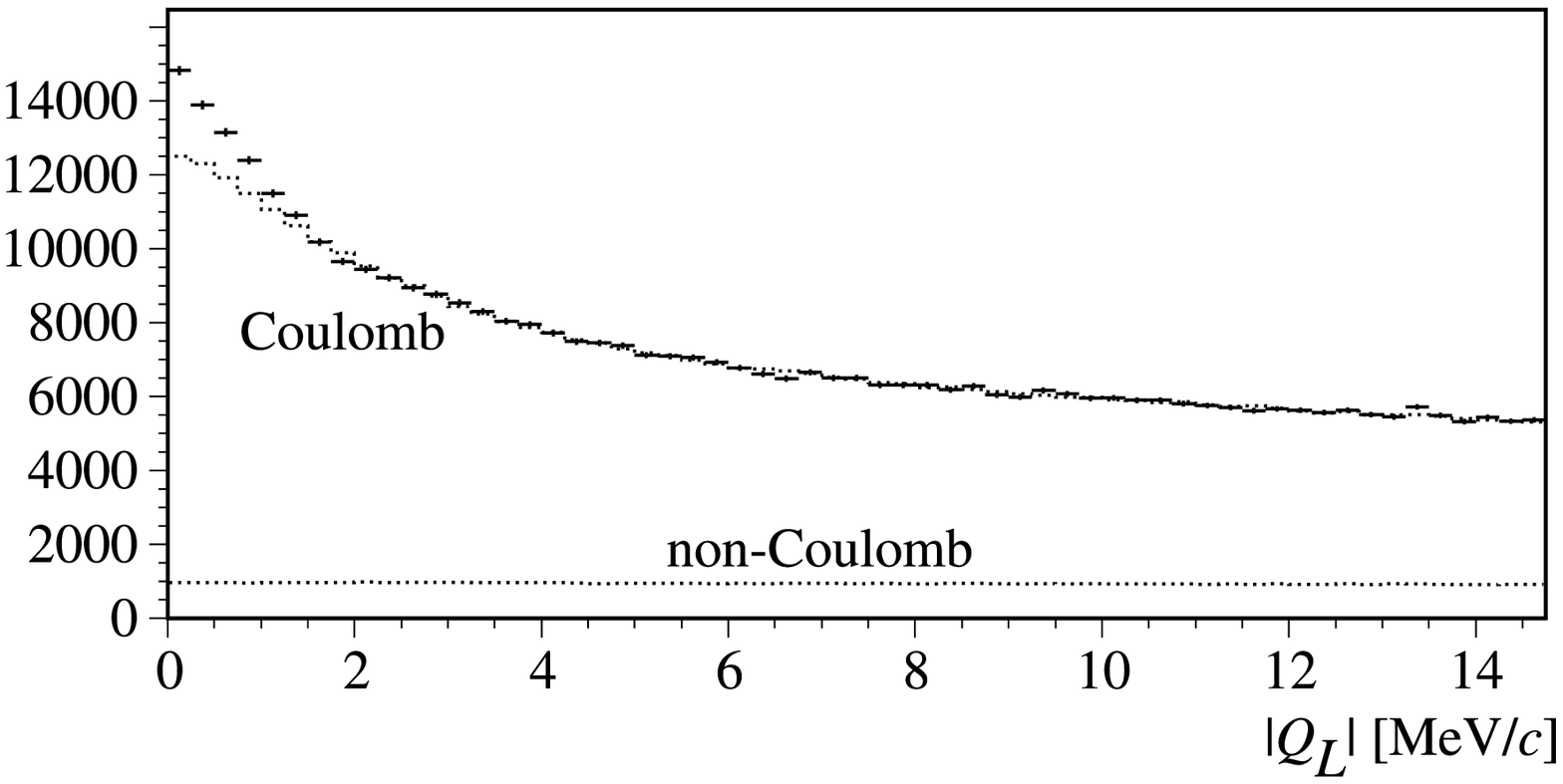}\\
   \includegraphics[width=0.47\textwidth]{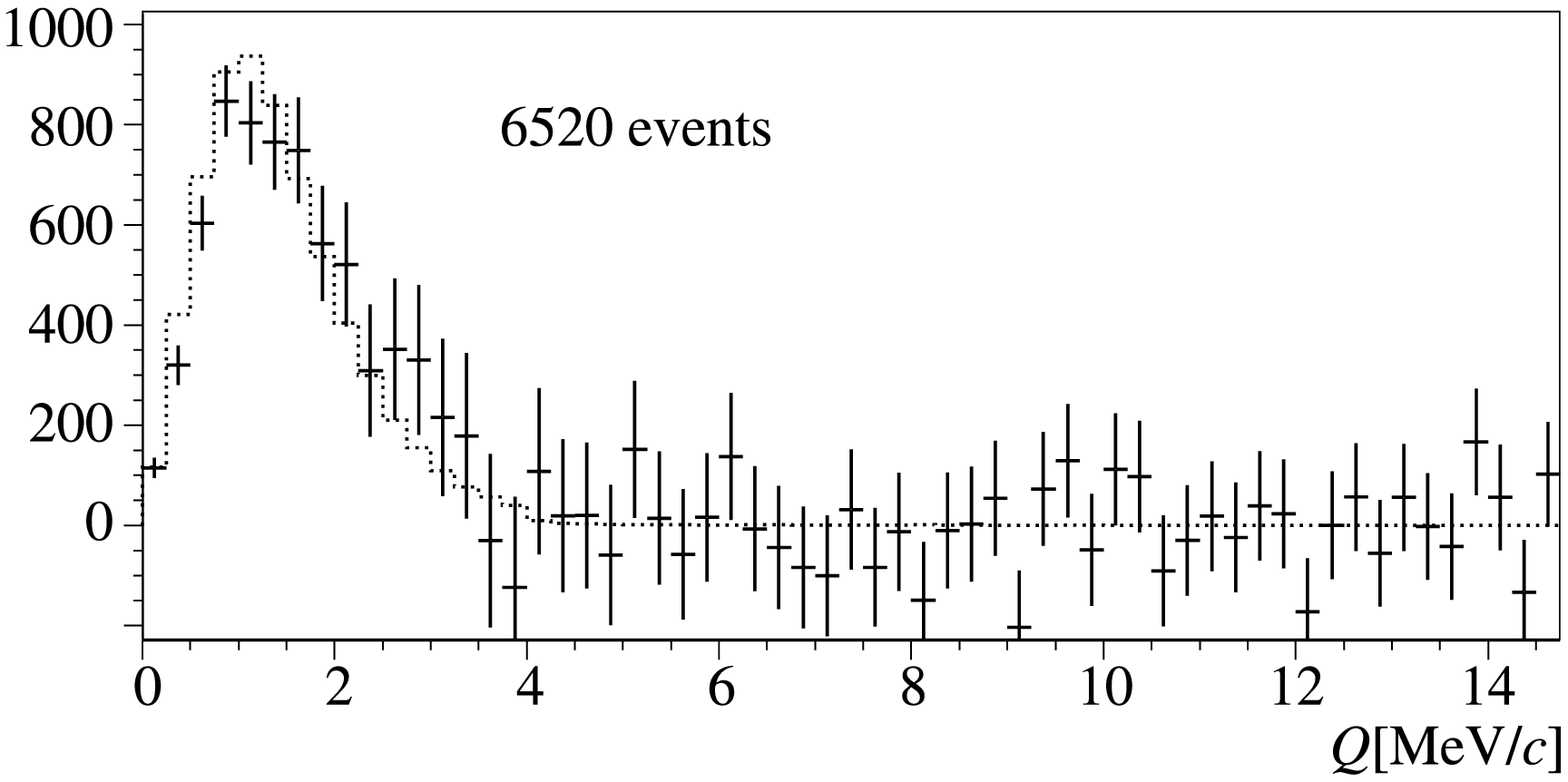}&
   \includegraphics[width=0.47\textwidth]{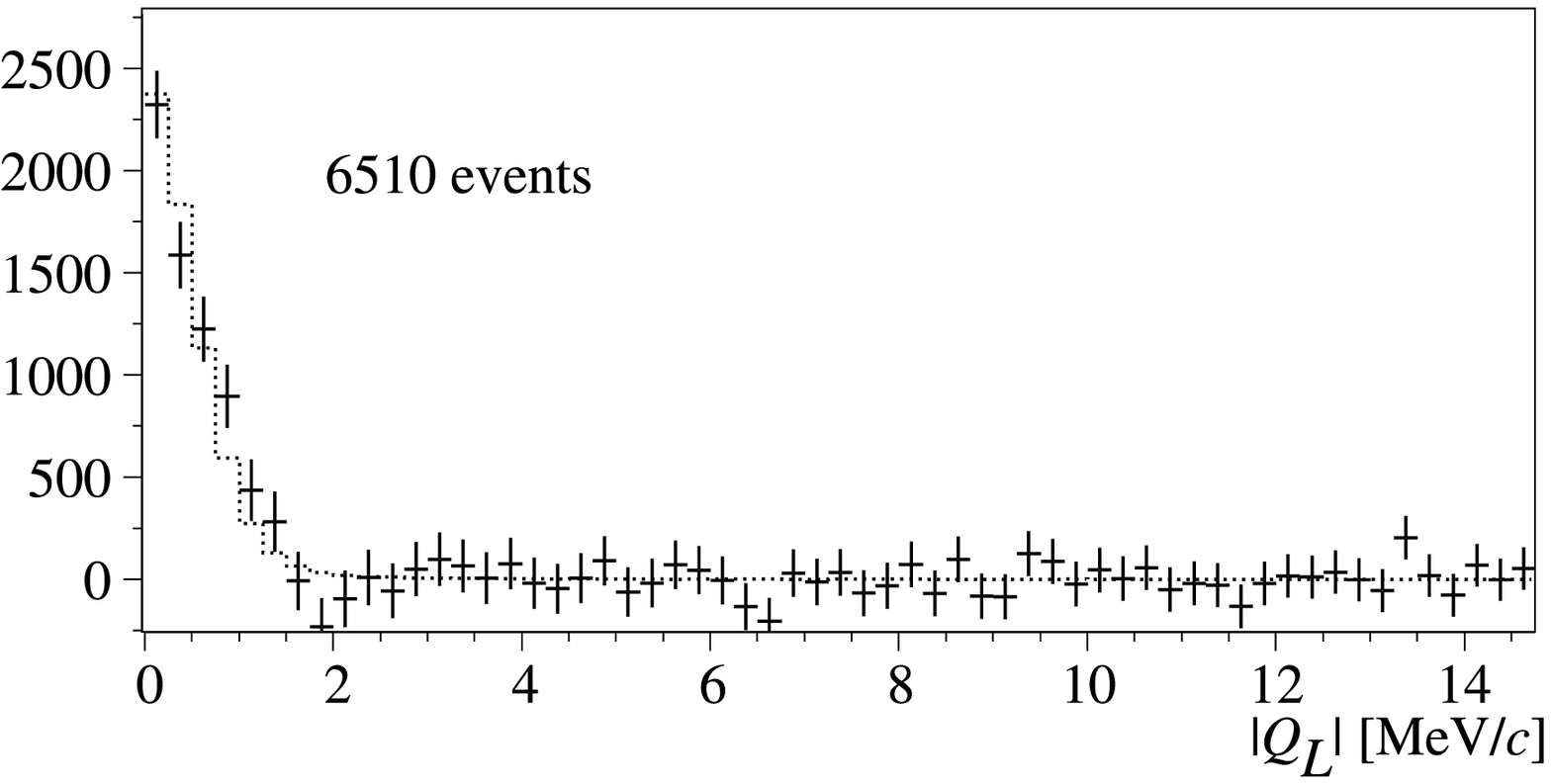}
\end{tabular}
\caption{Top: Experimental $Q$ and $Q_{L}$
  distributions after subtraction of the time uncorrelated background,
  approximated with Monte Carlo backgrounds (dashed lines). Bottom:
  Residuals after background subtraction.  The dashed lines represent
  the expected atomic signal shape. Notice that the signal strengths
  in $Q$ and $\mathrm{Q_{L}}$ are about the same.  The bin-width is
  0.25\,MeV/$c$.  }
\label{fig:BS-fit}
\end{figure}

The analysis using Monte Carlo backgrounds and no vertex fit is
summarized in Fig.\,\ref{fig:BS-fit}.  The background composition is
indicated, and the excess at low $Q$ and $Q_{L}$ is clearly seen.
Subtraction provides the residuals, also shown in
Fig.\,\ref{fig:BS-fit}, which represent the atomic signals. As expected
from Fig.\,\ref{fig:At}, the signal is narrower in $Q_{L}$ than in $Q$.
The simulated shape is in agreement with the data. The signal strength
has to be the same in $Q$ and $Q_{L}$ if the background is properly
reconstructed. The observed difference demonstrates that the
backgrounds are consistent at the per mille level.
The fact that outside of the signal region the residuals are perfectly
zero demonstrates the correctness of the Monte Carlo simulation.

\begin{figure}[htb]
 \centering
\begin{tabular}{cc}
   \includegraphics[width=0.47\textwidth]{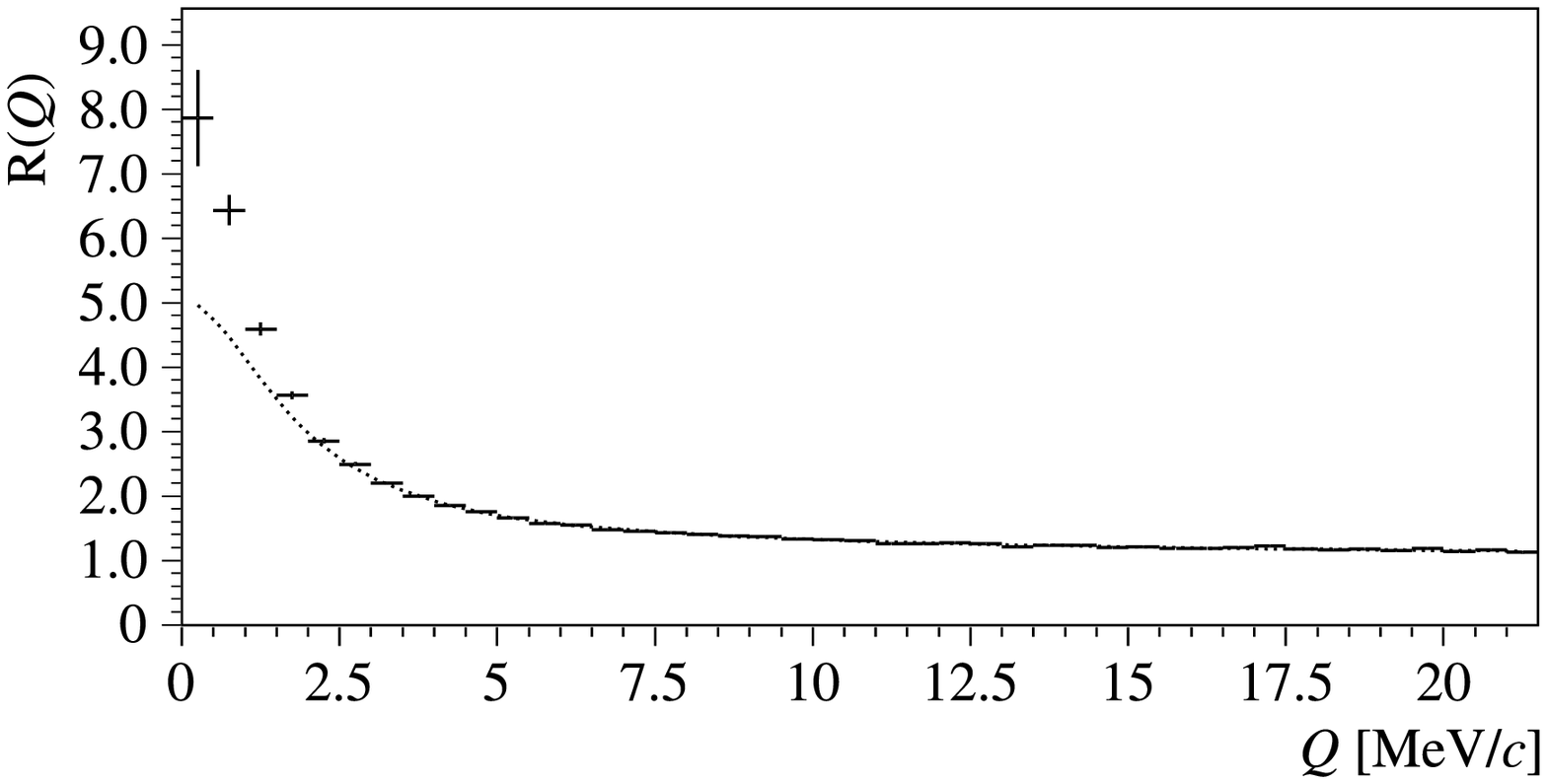}&
   \includegraphics[width=0.47\textwidth]{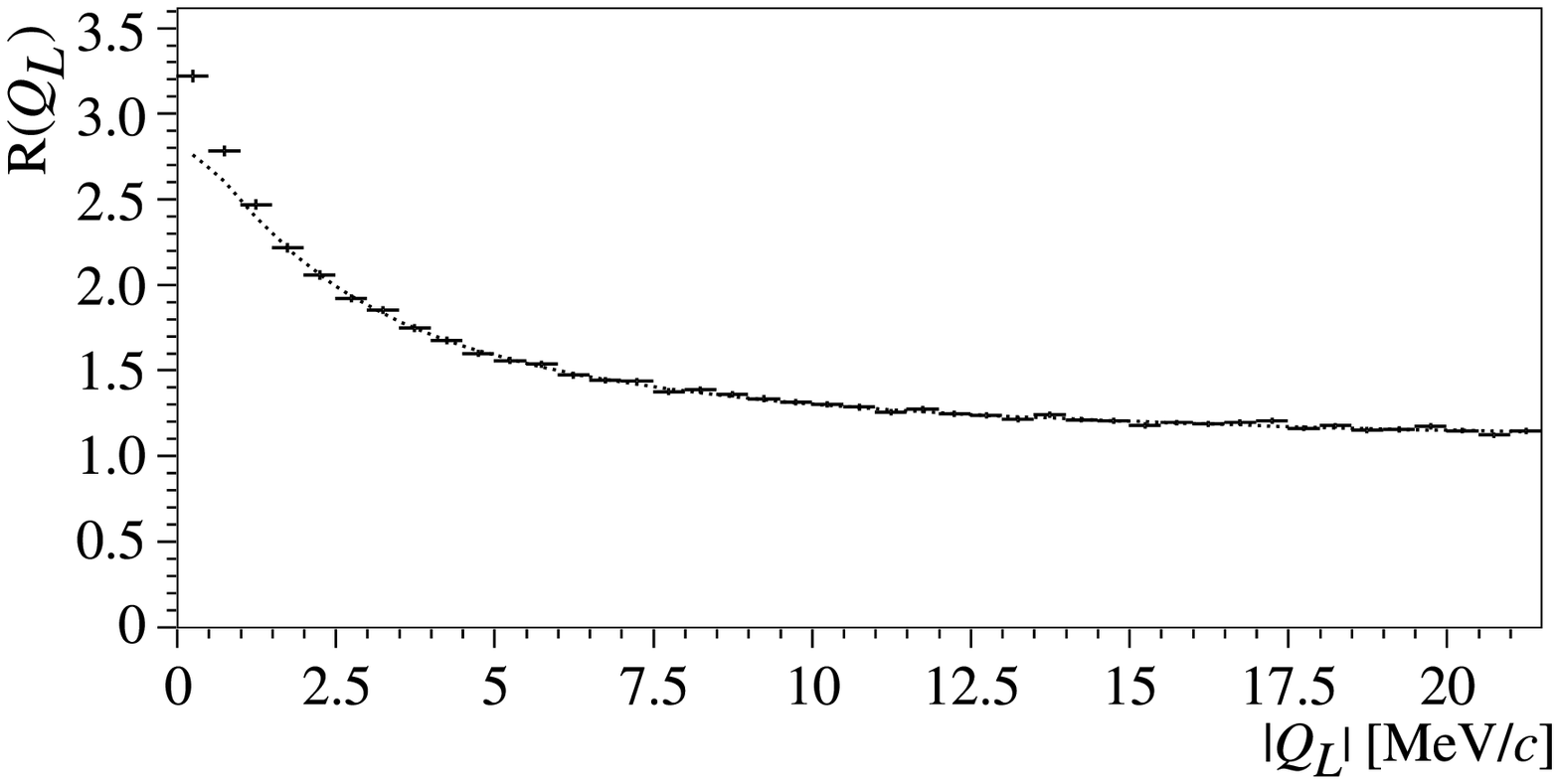}\\
   \includegraphics[width=0.47\textwidth]{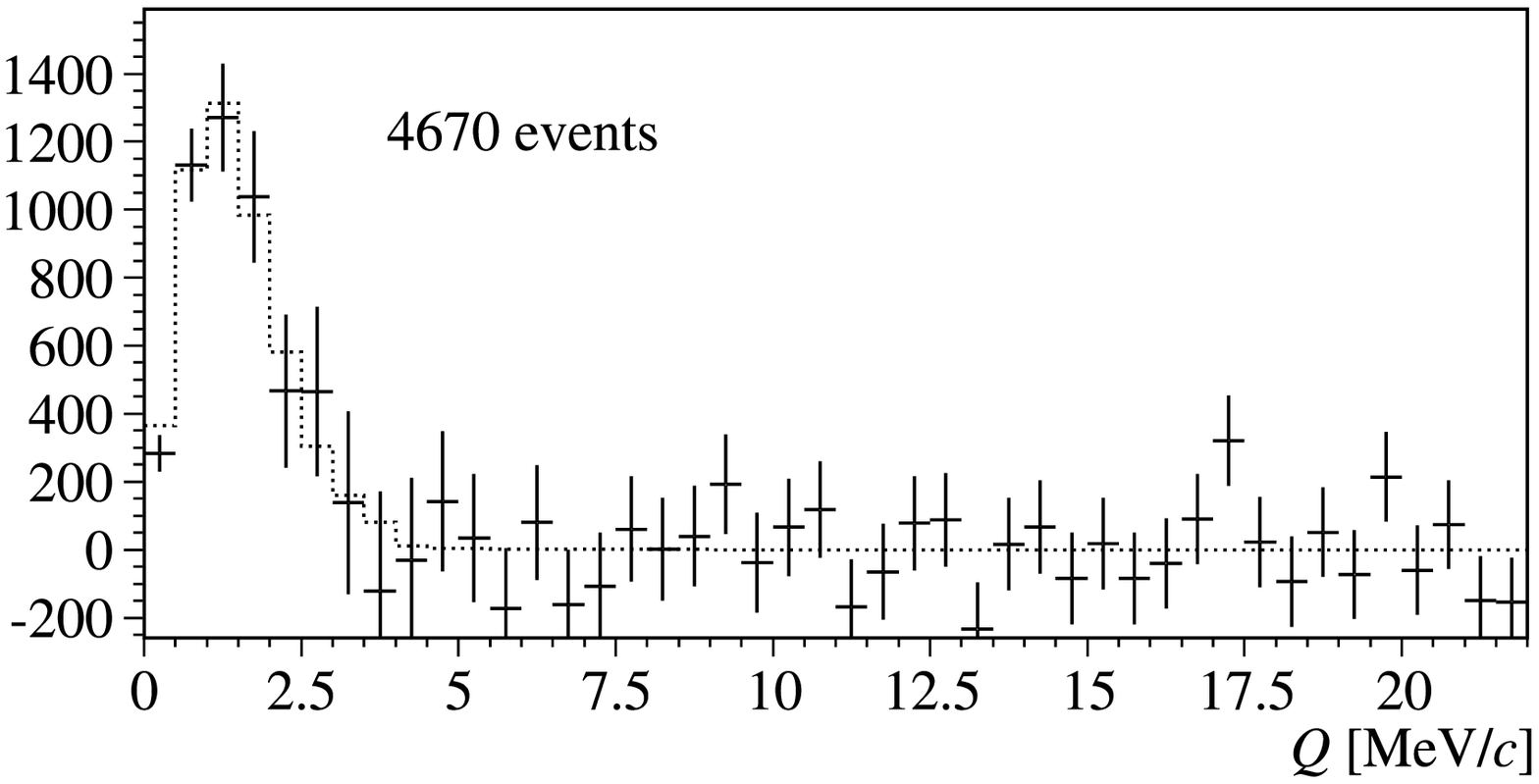}&
   \includegraphics[width=0.47\textwidth]{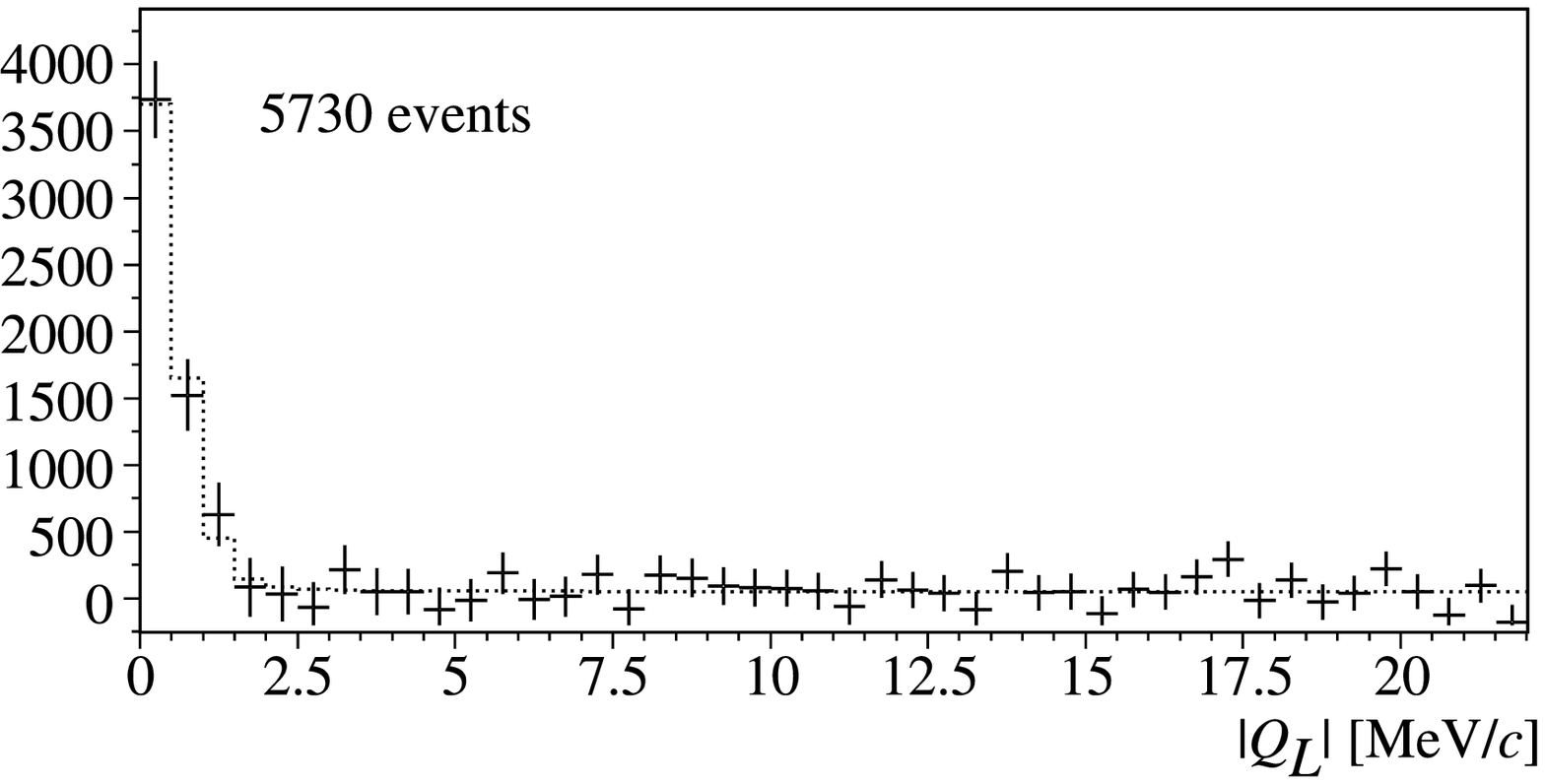}
\end{tabular}
\caption{Top: Experimental $Q$ and $Q_{L}$ correlation functions
(experimental points). The background correlation function is shown
as a smooth curve.  Bottom: Residuals after background subtraction. The
dashed lines  represent the expected atomic signal shape. The bin-width is
0.5\,MeV/$c$. }
\label{fig:Y-fit}
\end{figure}

The results of the analysis using the accidentals as a basis for
background modelling are shown in Fig.\,\ref{fig:Y-fit}. The clear
deviation of $R(Q)$ at low $Q$ from unity is due to the attractive
Coulomb interaction in the final state.  There is good agreement
between the experimental and fitted "correlation" function $R$ for $Q
> 2.5$\,MeV/$c$, whereas for $Q < 2.5$\,MeV/$c$ the experimental
correlation function (left hand side of Eq.\,\ref{eq:eqY}) shows
significantly higher values, due to the presence of atomic pairs. In
order to extract the number of atomic pairs the fit function (right
hand side of Eq.\,\ref{eq:eqY}) has been multiplied by the accidental
distribution $\Phi(Q)$. The subtraction of the latter from the full
measured spectrum leads to the atomic pair signal in
Fig.\,\ref{fig:Y-fit}.

Based on the same data sample of preselected events, the analysis using
Monte Carlo backgrounds and no vertex fit was repeated allowing for
three hits instead of only two in the search window (cf. section
\ref{selection}) and selecting the hit closest to the search reference
trajectory. This yields about 25\% more events in the signal, but 43\%
more in total (see Table\,\ref{tab:res}), thus the signal/background
quality diminishes.

An analysis fully independent on the upstream detectors was done using
only information from the downstream detectors. Events were selected
from raw data along with a looser cut in $Q_{T}$. The particles were
assumed to be emitted from the center of the beam spot at the target.
For this analysis only the $Q_{L}$ distribution shows a signal,
because multiple scattering in the upstream detectors does not distort
$Q_{L}$ so much but destroys $Q_{T}$ and hence $Q$. With respect to
the standard selection the signal increases by about 40-50\%. The
total number of events is, however, almost 3.5 times larger (see
Table\,\ref{tab:res}).  The background increases with respect to the
standard selection by a factor of two because of the loose cut on
$Q_{T}\leq 6$MeV/$c$. Moreover, signal and background increase because
\begin{itemize}
\item no cut on the multiplicity two in the SFD search window is
  needed (gain $\geq 25$\%).
\item no detector response of the SFD is needed (gain due to single
  track efficiency of 95\% for two tracks in two planes $\geq 23$\%.).
\end{itemize}

\begin{table}[ht]
   \caption{\label{tab:res} Reconstructed atomic pair events for $Q$ 
and $Q_{L}$ using full tracking and Monte Carlo background (Full), no
vertex fit and Monte Carlo background (MC), vertex fit and background
reconstruction from accidentals (ACC), as "MC" but using up to 3 hits
instead of two in the SFD acceptance (MC-3 hits), and reconstruction
only with downstream detectors (Down). Full selected data sample used
for signal extraction (integral from 0 to 15\,MeV/$c$) for each method
(selected sample), and ratio of extracted signal and selected data
sample ($\frac{\mathrm{signal}\;Q_{L}}{\mathrm{selected\;sample}}$). }
\begin{indented}
\item[]\begin{tabular}{cccccc} \br &Full&MC & ACC & MC-3 hits & Down\\ 
    \mr
    $Q$ &$5526\pm 385$& $6520\pm 370$ & $4670\pm 930$ & 8230$\pm$440 &- \\
    $Q_{L}$ &$5322\pm 350$&$6510\pm 330$ & $5730\pm 580$ & 8050$\pm$380 & $9280\pm 970$\\
    selected sample &$\sim 429000$&$437060$ & $385870 $ & 624880 & 1503700\\
    $\frac{\mathrm{signal~ }Q_{L}}{\mathrm{selected ~sample}}$
    &1.24\% & 1.49\% & $1.48 \% $ & 1.29\% & 0.6\% \\ \br
\end{tabular}
\end{indented}
\end{table}

The results are summarized in Table\,\ref{tab:res}. The column "Full"
shows the results obtained with full tracking (cf. section
\ref{tracking}) and background from Monte Carlo. The column "MC" shows
the result from reconstruction using Monte Carlo backgrounds and no
vertex fit \cite{CS-thesis}. The column "ACC" shows the result of
reconstruction using the measured accidentals and vertex fit.  The
column "MC-3 hits" is analogous to column "MC" but allowing for 3 hits
in the SFD planes \cite{CS-thesis} instead of two (cf. section
\ref{selection}). The column "Down" shows the result of using only the
downstream detectors and the background reconstruction based on
accidentals.  The integral (0 to 15\,MeV/$c$) number of events
retained after the specific selection cuts of each analysis methods is
also given.  The integral number of measured accidentals (needed for
background reconstruction for col.  "ACC") is about the same as the
number of prompt events.  The last row of Table\,\ref{tab:res}
shows the signal fraction extracted from the selected data sample.

In the case of full tracking (col. "Full" in Table\,\ref{tab:res}) the
integrated luminosity was (for the Ni-runs in 2001) only 79\% of the
preselected data due to the availability of the MSGC detector.
Moreover, these data were not subject to a cut on 2 hits in the SFD
(section \ref{selection}) and thus should be compared with column
"MC-3 hits" of Table\,\ref{tab:res}.

First we observe that the number of events in the full sample depends
on the details of the cut procedures and selection methods. Using
downstream detectors only (col."down" of Table\,\ref{tab:res}) results
in a background much larger than the signal increase.

The consistency of the signal is satisfactory as can be seen from
Table\,\ref{tab:res}. The signal fractions of the selected data sample
are the same for the Monte Carlo based method and the method based on
accidentals (Table\,\ref{tab:res}, columns "MC" and "ACC"), and very
similar for full tracking and MC-3 hits (Table\,\ref{tab:res},
columns "Full" and "MC-3 hits").  The large errors of column "ACC" of
Table\,\ref{tab:res} are due to the limited statistics of the measured
accidentals.  The difference in signal strength for $Q$ and $Q_{L}$ is
due to slightly different fit regions for $Q$ ($Q_T < 4$\,MeV/$c$,
$|Q_L| < 22$\,MeV/$c$, $Q > 4$\,MeV/$c$) and $Q_L$ ($Q_T <
4$\,MeV/$c$, 2\,MeV/$c<|Q_L| < 22$\,MeV/$c$) and for signal searching
($Q < 4$\,MeV/$c$ and $|Q_L| < 2$\,MeV/$c$, respectively). Thus, 
the backgrounds in $Q$ and $Q_L$ are not identical and 
fluctuations of the background differences in the prompt data and in
the measured accidental data lead to the (accidentally large) difference
in the signal. The same argument holds for the Monte Carlo based
method.  However, there the statistical fluctuations of the background
are negligible as compared to the measured data.  Allowing for three
hits instead of two in the SFDs yields about the same number of events
in the signal as were found without using any upstream detector.  This
indicates that the main source of loss in signal is inefficiency in
the upstream detectors.


\section{Conclusion}
For the first time a large statistics sample of $\pi^{+}\pi^{-}$ pairs
from atom break-up has been detected. Independent tracking procedures
and complementary background reconstruction strategies lead to
compatible results.  The background as obtained with Monte Carlo
methods is in excellent agreement with the data.  We conclude that
severe systematic errors may be excluded when extracting the atomic
pair signal.  The statistical accuracy of the signal allows for an
estimated statistical error on the lifetime of the $A_{2\pi}$ atom of
about 15\% \cite{CS-thesis}. In this paper we have analysed only
part of our data. In view of the full statistics accumulated by the
experiment so far and with systematic errors estimated to be smaller
than the statistical ones \cite{CS-thesis} the goal of the experiment
to achieve a 10\% accuracy for the lifetime of the $A_{2\pi}$ atom is
in reach.


\ack

We are indebted to the CERN PS crew for providing a beam of excellent
quality.  Some of us acknowledge support from ETC$^*$, Trento.  This
work was supported by CERN, the Grant Agency of the Czech Republic,
grant No. 202/01/0779, the Greek General Secretariat of Research and
Technology (Greece), the University of Ioannina Research Committee
(Greece), the IN2P3 (France), the Istituto Nazionale di Fisica
Nucleare (Italy), the Japan Society for the Promotion of Science
(JSPS), Grant-in-Aid for Scientific Research No. 07454056, 08044098,
09640376, 11440082, 11694099, 12440069, 14340079 and 15340205.  JINR
Dubna, contract No. 08626319/ 96682-72, the Ministery of Industry,
Science and Technologies of the Russian Federation and the Russian
Foundation for Basic Research (Russia), under project 01-02-17756, the
Swiss National Science Foundation, the Ministerio de Ciencia y
Tecnologia (Spain), under projects AEN96-1671 and AEN99-0488, the
PGIDT of Xunta de Galicia (Spain).


\section*{References}

\begin{thebibliography}{99} 
\bibitem{Adeva95} Adeva~B \textit{et al.}, \textit{DIRAC proposal},
  CERN/SPSLC 95-1, SPSLC/P 284 (1995)

\bibitem{Deser54} Deser~S \textit{et al.}, \textit{Phys. Rev.} \textbf{96}
  (1954) 774
  
\bibitem{Uretsky61} Uretsky~J and Palfrey~J, \textit{Phys. Rev.} \textbf{121}
  (1961) 1798
  
\bibitem{Bilenky69} Bilenky~S~M \textit{et al.}, \textit{Yad. Fiz.}
  10 (1969) 812; (\textit{Sov. J. Nucl. Phys.} \textbf{10} (1969) 469)
  
\bibitem{Jallouli98} Jallouli~H and Sazdjian~H, \textit{Phys. Rev.}
  \textbf{D58} (1998) 014011; Erratum: \textit{ibid}. \textbf{D58} (1998) 099901
  
\bibitem{Ivanov98} Ivanov~M~A \textit{et al.} \textit{Phys. Rev.} \textbf{D58}
  (1998) 094024
  
\bibitem{Gasser01} Gasser~J \textit{et al.}, \textit{Phys. Rev.} \textbf{D64}
  (2001) 016008; hep-ph/0103157

\bibitem{Gashi02} Gashi~A \textit{et al.}, \textit{Nucl. Phys.} \textbf{A699} (2002) 732

\bibitem{Weinb79} Weinberg~S, \textit{Physica} \textbf{A96} (1979) 327;
Gasser~J and Leutwyler~H, \textit{Phys. Lett.} \textbf{B125} (1983) 325 and
\textit{Nucl. Phys.} \textbf{B250} (1985) 465, 517, 539

\bibitem{Colan01NP} Colangelo~G, Gasser~J and Leutwyler~H,
  \textit{Nucl. Phys.}  \textbf{B603} (2001) 125

  
\bibitem{Colan01PRL} Colangelo~G, Gasser~J and Leutwyler~H,
  \textit{Phys. Rev. Lett.} \textbf{86} (2001) 5008
  
\bibitem{Pislak01} Pislak~S \textit{et al.}, \textit{Phys. Rev. Lett.}
  \textbf{87} (2001) 221801
  
\bibitem{Knecht95} Knecht~M \textit{et al.}, \textit{Nucl. Phys.} \textbf{B457}
  (1995) 513
  
\bibitem{Nem85} Nemenov~L~L, \textit{Yad. Fiz.} \textbf{41} (1985) 980;
  (\textit{Sov. J.  Nucl. Phys.} \textbf{41} (1985) 629)
  
\bibitem{Gorch96} Gorchakov~O~E \textit{et al.}, \textit{Yad. Fiz.} \textbf{59}
  (1996) 2015; (\textit{Phys. At. Nucl.} \textbf{59} (1996) 1942)
  
\bibitem{Gorch00} Gorchakov~O~E \textit{et al.}, \textit{Yad. Fiz.} \textbf{63}
  (2000) 1936; (\textit{Phys. At. Nucl.} \textbf{63} (2000) 1847)
  
\bibitem{Schum02} Schumann~M \textit{et al.}, \textit{J. Phys. B} \textbf{35}
  (2002) 2683.
  
\bibitem{Afan96} Afanasyev~L~G and Tarasov~A~V, \textit{Yad. Fiz.} \textbf{59}
  (1996) 2212, (\textit{Phys. At. Nucl.} \textbf{59} (1996) 2130)
  
\bibitem{Dulian83} Dulian~L~S and Kotsinian~A~M, \textit{Yad. Fiz.} \textbf{37}
  (1983) 137; (\textit{Sov. J. Nucl. Phys.} \textbf{37} (1983) 78)
  
\bibitem{Mrowc} Mr\'owczy\'nski~S, \textit{Phys. Rev.} \textbf{A33} (1986) 1549;
  Mr\'owczy\'nski~S, \textit{Phys. Rev.} \textbf{D36} (1987) 1520;
  Denisenko~K~G and Mr\'owczy\'nski~S, \textit{ibid}. \textbf{D36} (1987) 1529
  
\bibitem{Halab99} Halabuka~Z \textit{et al.}, \textit{Nucl. Phys.}
  \textbf{B554} (1999) 86
  
\bibitem{Taras91} Tarasov~A~V and Khristova~I~U, JINR-P2-91-10, Dubna
  1991
  
\bibitem{Voskr98} Voskresenskaya~O~O, Gevorkyan~S~R and Tarasov~A~V,
  \textit{Phys. At. Nucl.} \textbf{61} (1998) 1517

\bibitem{Afan99} Afanasyev~L, Tarasov~A and Voskresenskaya~O,
  \textit{J.~Phys.~G} \textbf{25} (1999) \textbf{B7}
  
\bibitem{Ivanov99gl} Ivanov~D~Yu, Szymanowski~L, \textit{Eur. Phys.
    J.} \textbf{A5} (1999) 117
  
\bibitem{Heim00} Heim~T~A \textit{et al.}, \textit{J~Phys~B} \textbf{33} (2000)
  3583
  
\bibitem{Heim01} Heim~T~A \textit{et al.}, \textit{J~Phys~B} \textbf{34} (2001)
  3763

\bibitem{Afan02} Afanasyev~L, Tarasov~A and Voskresenskaya~O,
  \textit{Phys. Rev.} D 65 (2002) 096001; hep-ph/0109208

  
\bibitem{Santa03} Santamarina~C \textit{et al.}, \textit{J~Phys~B} \textbf{36}
  (2003) 4273

\bibitem{Afan93} Afanasyev~L~G \textit{et al.}, \textit{Phys. Lett.}
  \textbf{B308} (1993) 200
  
\bibitem{Afan94} Afanasyev~L~G \textit{et al.}, \textit{Phys. Lett.}
  \textbf{B338} (1994) 478
  
\bibitem{Afana96} Afanasyev~L~G \textit{et al.}, Communication JINR
  P1-97-306, Dubna, 1997
  
\bibitem{setup} Adeva~B \textit{et al.}, \textit{Nucl. Instr. Meth.}
  \textbf{A515} (2003) 467

\bibitem{PS/CA/Note97-16} Ferrando~O and Hemery~J-Y, CERN-PS
  Division, PS/ CA/ Note97-16

\bibitem{note02-01}  Lanaro~A, \textit{DIRAC note} 2002-01,
  http://dirac.web.cern.ch/DIRAC/i\_notes.html
  
\bibitem{PDG2002} Hagiwara~K \textit{et al.} (PDG), \textit{Phys.
    Rev.} \textbf{D66} (2002) 010001

  
\bibitem{santiago-lambda} Adeva~B, Romero~A and Vazquez~Doce~O,
  DIRAC note 2004-05,\\
  http://dirac.web.cern.ch/DIRAC/i\_notes.html

  
\bibitem{dirac-resolution} Gortchakov~O and Santamarina~C,
  \textit{DIRAC note} 2004-01,\\
  http://dirac.web.cern.ch/DIRAC/i\_notes.html
  
\bibitem{ADEV02} Adeva~B \textit{et al.}, \textit{Nucl. Instr. Meth.}
  \textbf{A491} (2002) 41

 \bibitem{multilevel} Afanasyev~L \textit{et al.}
   \textit{Nucl. Instr. Meth.} \textbf{A491} (2002) 376

\bibitem{note03-04} Vlachos~S, DIRAC note 2003-04, 
  http://dirac.web.cern.ch/DIRAC/i\_notes.html

\bibitem{1stlevel} Afanasyev~L \textit{et al.},
  \textit{Nucl. Instr. Meth.} \textbf{A479} (2002) 407

\bibitem{T3} Gallas~M,
\textit{Nucl. Instr. Meth.} \textbf{A481} (2002) 222

\bibitem{DNA} Kokkas~P \textit{et al.}, 
  \textit{Nucl. Instr. Meth.} \textbf{A471} (2001) 358

\bibitem{note00-13} Vlachos~S, DIRAC note 2000-13,
  http://dirac.web.cern.ch/DIRAC/i\_notes.html
  
\bibitem{DAQ} Olshevsky~V~G, Trusov~S~V, \textit{Nucl. Instr. Meth.}
  \textbf{A469} (2001) 216
  
\bibitem{readout} Karpukhin~V \textit{et al.}, \textit{Nucl. Instr.
    Meth.} \textbf{A512} (2003) 578
  
\bibitem{deadtime} Kulikov~A, Zhabitsky~M, \textit{Nucl. Instr.
    Meth.} \textbf{A527} (2004) 591

  
\bibitem{GEANT-DIRAC} \textit{GEANT3 for DIRAC}, version 2.63,
  unpublished, http://dirac.web.cern.ch/DIRAC/ 
  
\bibitem{ARIANE-DIRAC} \textit{Analysis software package for DIRAC},
  unpublished
  
\bibitem{KALMAN} Maybeck~P, \textit{Stochastic Models, Estimation and
    Control}, Vol. 1. Academic Press, New York (1979)

\bibitem{BSTRACK} Ch. Schuetz~P and Tauscher~L, DIRAC note 02-01, 
Ch. Schuetz~P and Tauscher~L, DIRAC note 03-06,
http://dirac.web.cern.ch/DIRAC/i\_notes.html

\bibitem{Pentia} Pentia~M \textit{et al.}, \textit{Nucl. Instr. Meth.}
  \textbf{A369} (1996) 101; Pentia~M and Constantinescu~S, DIRAC note
  01-04, http://dirac.web.cern.ch/DIRAC/i\_notes.html

\bibitem{S-TRACK} Adeva~B, Romero~A and Vazquez Doce~O, DIRAC note
2003-08,\\http://dirac.web.cern.ch/DIRAC/i\_notes.html

\bibitem{kokkas-lambda}Kokkas~P, DIRAC note 2004-04, 
  http://dirac.web.cern.ch/DIRAC/i\_notes.html

\bibitem{note01-02} Brekhovskikh~V and Gallas~M~V, DIRAC note
2001-02,\\ http://dirac.web.cern.ch/DIRAC/i\_notes.html

\bibitem{note03-09} Santamarina~C and Ch. Schuetz~P, DIRAC note
2003-09,\\ http://dirac.web.cern.ch/DIRAC/i\_notes.html

\bibitem{SAKH48} Sakharov~A~D, \textit{Zh. Eksp. Teor. Fiz.} \textbf{18} (1948)
  631
  
\bibitem{ll82} Lednicky~R, Lyuboshitz~V~L, \textit{Yad. Fiz.} \textbf{35}
  (1982) 1316 (\textit{Sov. J. Nucl. Phys.} \textbf{35} (1982) 770)

\bibitem{note01-01} Lanaro~A, DIRAC note 2001-01,
  http://dirac.web.cern.ch/DIRAC/i\_notes.html
  
\bibitem{CS-thesis} Ch. Schuetz~P, Thesis, Basel University, March
  2004,\\http://cdsweb.cern.ch/search.py?recid=732756

  
\end{thebibliography}
\end{document}